\documentclass[english,aps,twocolumn,=superscriptaddress,prb,eqsecnum,notitlepage,nofootinbib]{revtex4}
\usepackage{babel}
\usepackage{amsmath}
\usepackage{amssymb}
\usepackage{graphicx}
\usepackage{epsfig,natbib}
\usepackage{mathrsfs} 
\usepackage{color}

\usepackage{hyperref}
\hypersetup{
        colorlinks=true,
}

\newcommand{\be}{\begin{equation}}
\newcommand{\ee}{\end{equation}}
\newcommand{\bea}{\begin{eqnarray}}
\newcommand{\eea}{\end{eqnarray}}
\newcommand{\ket}{\rangle}
\newcommand{\bra}{\langle}

\begin{document}
\title{Majorana-Hubbard model on the square lattice}
\author{Ian Affleck}
\affiliation{Department of Physics and Astronomy and Stewart Blusson Quantum Matter Institute, University of British Columbia, 
Vancouver, B.C., Canada, V6T1Z1}
\author{Armin Rahmani}
\affiliation{Department of Physics and Astronomy and Advanced Materials Science and Engineering Center, Western Washington University, Bellingham, Washington 98225, USA}
\author{Dmitry Pikulin}
\affiliation{Station Q, Microsoft Research, Santa Barbara, California 93106-6105, USA}

\begin{abstract}
We study a tight-binding model of interacting Majorana (Hermitian) modes on a square lattice. The model may have  an 
experimental realization in a superconducting-film--topological-insulator heterostructure in a magnetic field. We find a rich phase diagram, as a function of 
interaction strength, including an emergent superfluid phase with spontaneous breaking of an emergent $U(1)$ symmetry, separated 
by a supersymmetric transition from a gapless normal phase. 
\end{abstract}
\date{\today} 
\maketitle
\section{Introduction, model, and Phase diagram}\label{sec:intro}
The discovery of topological materials~[\onlinecite{Konig2007, Zhang2009, Xia2009, Mourik2012}] has led to great interest in Majorana modes (MM)~[\onlinecite{Alicea2012,Beenakker2013a}], which are promising candidates for topological quantum computing~[\onlinecite{Nayak2008, Hyart2013, Aasen2016, Karzig2016}]. The MM's are predicted to appear in various situations at topological defects and boundaries of topological superconductors~[\onlinecite{Kitaev2001, Fu2008, Lutchyn2010, Oreg2010}]. 
In addition to theoretical proposals (and subsequent experimental progress) for realizing a separated localized MM ~[\onlinecite{Fu2008, Lutchyn2010, Oreg2010, Mourik2012, Albrecht2016}], in certain situations, in both one and two dimensions, a finite density of MM's is expected ~[\onlinecite{Fu2008, Zhou2013, Biswas2013, Chiu2015, Liu2015, Pikulin2015}]. 
The effects of interactions between MM's in such setups is a relatively unexplored field~[\onlinecite{Hermanns2014,Chiu2015, Pikulin2015, Rahmani2015, Rahmani2015a, Milsted2015, Witczak2016, Ware2016}].  
Interaction effects have also been studied~[\onlinecite{Gangadharaiah2011,Lobos2012}] in the Kitaev model~[\onlinecite{Kitaev2001}] in which two localized Majorana modes appear at the ends of a chain. 
The corresponding Hamiltonians
are Hubbard-like (with Majorana fermions serving as Hermitian counterparts of electrons), but necessarily have interactions spanning at least four lattice sites, since the square of a 
Majorana operator is a constant. 

The simplest such model, defined on a chain, was shown to have a rich phase 
diagram, with four different phases~[\onlinecite{Rahmani2015, Rahmani2015a,Milsted2015}]. The continuum limit involved a single species of massless relativistic Majorana fermions 
and interactions that are irrelevant when weak enough, since interactions necessarily contain 
derivatives of the Majorana fields. At strong enough coupling, MM's like to pair up 
on neighboring sites to form ordinary Dirac (non-Hermitian) fermions, leading to spontaneously broken 
translation symmetry due to the dimerization. The transition into this broken symmetry phase for attractive interactions was shown to be described by 
the tricritical Ising model, which exhibits supersymmetry. A phase with emergent $U(1)$ symmetry was also found 
at intermediate strength repulsive interactions, corresponding to a Luttinger liquid plus a relativistic 
Majorana fermion. Stronger repulsive interactions again led to Majorana fermions combining on neighboring 
sites to form Dirac fermions but in this case there is a further breaking of translational symmetry.

\begin{figure*}[th]
\centerline{\includegraphics[width=\linewidth]{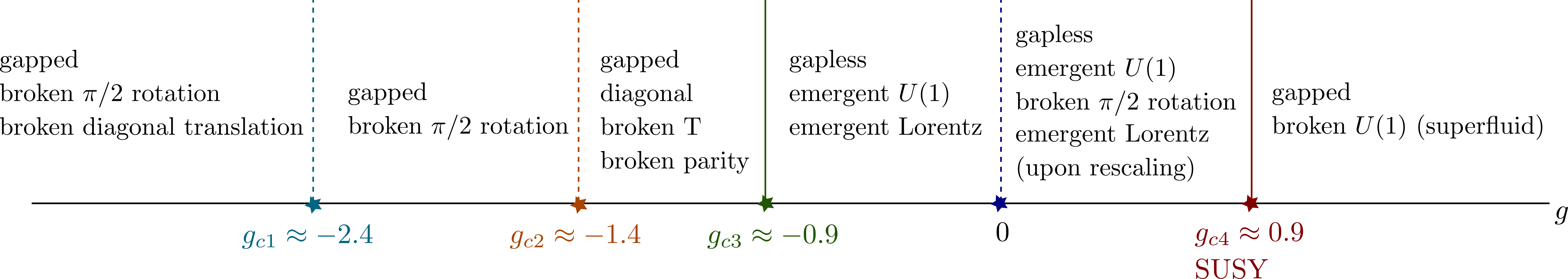}}
\caption{The mean-field phase diagram for $t=1$ and $t_2=0$ as a function of $g$. The  solid (dashed) lines represent second-order (first-order) transitions.  The broken symmetry states are sketched in Figs. (\ref{VF}, \ref{AVF}, and \ref{DgN}).}
\label{fig:phase_diagram}
\end{figure*}

In this paper, we study a two-dimensional model of interacting Majorana modes, motivated by a possible experimental realization in a superconducting 
thin film placed in a perpendicular magnetic field on top of a strong three-dimensional topological insulator. For simplicity, we consider the case 
of a square lattice of vortices, with each of them containing a Majorana mode. Requiring one superconducting flux quantum at 
each vortex determines the signs of the hopping terms to be ~[\onlinecite{Grosfeld2006, Liu2015}]
\be \label{eq:H0}H_0=it\sum_{m,n}\gamma_{m,n}[(-1)^n\gamma_{m+1,n}+\gamma_{m,n+1}],\ee
up to a $\mathbb{Z}_2$ gauge transformation, where $n$ and $m$ are integers. (The sign alternation of the horizontal hopping term can be changed by a gauge transformation but sign 
alternation cannot be completely removed.) We include the shortest possible range interaction term, occurring on plaquettes:
\be H_{int}=g\sum_{m,n}\gamma_{m,n}\gamma_{m+1,n}\gamma_{m+1,n+1}\gamma_{m,n+1}.\ee
We expect~[\onlinecite{Chiu2015}]  the actual interactions between Majorana modes to exhibit exponential decay; this short-range interaction, analogous to the Hubbard interaction for complex fermions, is a convenient simplification. 
\footnote{Short-range Majorana-Majorana interactions may also occur in He$^3$~[\onlinecite{Park2015}]}
The sign of $g$ is linked to the gauge choice in $H_0$ and the sign of the underlying physical interaction.
We also  discuss the effects of a second-neighbor hopping term chosen to be consistent with the flux:
\be H_{2}=it_2  \sum_{m,n,s,s'}\gamma_{m,2n}\gamma_{m+s,2n+s'}\label{eq:t2}\ee
where $s$ and $s'$ are summed over $\pm 1$. 
The Majorana operators are Hermitian, 
\be \gamma_{m,n}=\gamma_{m,n}^\dagger \ee
and obey the anti-commutation relations:
\be \{\gamma_{m,n},\gamma_{m',n'}\}=2\delta_{m,m'}\delta_{n,n'},\ee
implying $\gamma_{m,n}^2=1$. 
Note that this model has no conserved particle number so no chemical potential can be introduced. The sign of $t$ 
can be changed by sign redefinitions of the $\gamma$'s; we choose $t>0$. There is thus only one dimensionless parameter, 
$g/t$, in the model. An accurate numerical  treatment of this model is an enormous challenge, similar to the Hubbard model, 
and we do not attempt it here. 
Instead, we tackle the model with a combination of field theory, renormalization group 
and mean-field approaches. A promising numerical approach may be density-matrix renormalization group calculations 
on ladders [\onlinecite{Zhu2016}]; we will present results on this in a later paper. 
Note that this model is fundamentally different and much more challenging to study 
than other lattice models involving two copies of Majorana fermions, which are 
actually equivalent to ordinary complex fermion models, and have a conserved particle number~[\onlinecite{Hayata2017}].  Unlike the model of Refs.~[\onlinecite{Xiang2015,Hayata2017}], our model does not appear to be amenable to Quantum Monte Carlo simulations.
As in the 1D works~[\onlinecite{Rahmani2015, Rahmani2015a}], we study the model for 
both signs of the coupling constant. It is 
not completely clear which sign of the interactions might occur in experiments due to the effectively 
attractive interactions in a superconductor. $g>0$ corresponds to attractive interactions, as we will see. 

 We find the following mean-field phase diagram, when $t_2=0$, 
sketched in Fig. \ref{fig:phase_diagram}:
\begin{itemize}
\item{For $0<g<g_{c4}$,  there is a gapless phase with broken $\pi/2$ rotation symmetry. This phase has an emergent conserved particle number, $U(1)$, symmetry. There is also an emergent Lorentz invariance, upon rescaling the $x$-coordinate.}
\item{For $g>g_{c4}$, the emergent $U(1)$ symmetry is spontaneously broken at the critical point $g_{c4}$, corresponding to an emergent superfluid phase.}
\item{For $g_{c3}<g<0$, there is a gapless phase with no broken symmetries. This phase has several emergent symmetries including Lorentz 
invariance and an emergent conserved particle number, $U(1)$, symmetry. The phase transition at $g=0$ is first order. }
\item{For $g_{c2}<g<g_{c3}$, there is a phase with spontaneously broken parity and time reversal. The phase transition is second order (first order) at $g_{c3}$ ($g_{c2}$).}
\item{For $g_{c1}<g<g_{c2}$, there is a phase with spontaneously broken $\pi /2$ spatial rotation symmetry. This phase has translation symmetry in the diagonal direction. }
\item{For $g<g_{c1}$, we have a phase that in addition to the $\pi /2$ spatial rotation, also breaks the diagonal translation symmetry. The transition at $g_{c1}$ is first order. }
\end{itemize} 
The broken symmetry phases occurring for $g<g_{c2}$ and $g>g_{c4}$ correspond to nearby Majorana modes combining in pairs to form Dirac fermions, in different patterns. 
The transition at $g_{c4}$ between gapless and emergent superfluid phases is in a ${\cal N}=2$ supersymmetric (SUSY) universality class 
[\onlinecite{Thomas2005,Lee2007,Ponte2014,Jian2015,Zerf2016}].  (This is distinct from other condensed matter realizations of ${\cal N}=1$ SUSY~[\onlinecite{Grover2014}].)
A nonzero $t_2$ produces a gap in the weak coupling phase, corresponding to a mass in the field theory. 
A transition to an emergent superfluid phase still occurs for large enough positive $g$ but we expect that it is now in the conventional 
$U(1)$ universality class, without supersymmetry. A nonzero $t_2$ explicitly breaks parity and time reversal, 
eliminating the transition at $g_{c3}$. The phases for $g<g_{c2}$ with spontaneously broken spatial rotation symmetry 
may still occur. 

The remainder of this paper is organized as follows.  In Sec. \ref{symm} we analyze the symmetries of the model. In Sec. \ref{NIS}
we solve the noninteracting model and discuss the topological classification of various phases. 
In Sec. \ref{FTS}, we derive the low-energy effective field theory and discuss the various emergent symmetries. In Sec. \ref{MFT_PD}, we  analyze
the effects of interactions in mean field theory. In Sec. \ref{sec:ft}, we discuss the broken 
symmetry phases using the low energy field theory, and the universality class of the continuous transitions at 
$g_{c3}$ and $g_{c4}$.  We close the paper in Sec. \ref{sec:sum} with a brief summary.

\section{Symmetries of the lattice Hamiltonian}
\label{symm}
The Hamiltonian has no continuous symmetries (in particular, no particle number conservation due to the Majorana nature 
of the fermions). However, it has 5 important discrete symmetries when $t_2=0$,
 which lead to continuous emergent symmetries in the low energy effective Hamiltonian. 

The model, with the second-neighbor hopping included, is  invariant under translation by 1 site in the $x$ or $y$ directions:
\bea \gamma_{m,n}&\to& \gamma_{m+1,n}\nonumber \\
\gamma_{m,n}&\to& (-1)^m\gamma_{m,n+1}
.\eea

  Without second neighbor hopping, the lattice Hamiltonian is invariant under the anti-unitary time reversal transformation, $T$:
  \bea \gamma_{m,n}&\to& (-1)^{m+n}\gamma_{m,n}\nonumber \\
  i&\to& -i.\eea
Time reversal symmetry is broken by the next-neighbor hopping term, $\propto t_2$. Since our model describes a system 
in a magnetic field, we expect time reversal symmetry to be broken so there is no reason to exclude $t_2$. However, 
we might hope that it is relatively small compared to $t_1$ so that time reversal is an approximate symmetry. 
  
The spatial parity symmetry, reflection in the $x$ axis, is
\be \gamma_{m,n}\to (-1)^m\gamma_{-m,n}.\ee
This can be seen to be a symmetry of the Hamiltonian only when second neighbor hopping, $t_2$ is excluded. The product 
of time reversal and parity {\it is} a symmetry even when second-neighbor hopping is included. 
  
  Finally, the Hamiltonian, including the second-neighbor hopping, is invariant under a $\pi /2$ spatial rotation:
\be \gamma_{m,n}\to s_{m,n}\cdot \gamma_{-n,m}\label{po2}\ee
where
\bea s_{m,n}&=&-1\ \ (m\ \hbox{even}\ \hbox{and}\   n\  \hbox{odd})\nonumber \\
&=&1\ \  (\hbox{otherwise}).\label{s}\eea
This is confirmed in Appendix A. 

\section{Solution of noninteracting model}
\label{NIS}
\subsection{The energy spectrum}
For the gauge chosen in \eqref{eq:H0} there are 2 sites per unit 
cell in the $y$-direction, with $n$ even and odd.  We relabel
\bea \gamma_{m,2n}&=&\gamma^e_{m,2n},\nonumber \\
\gamma_{m,2n+1}&=&\gamma^o_{m,2n}.\eea
Then we Fourier transform, imposing periodic boundary conditions:
\be \gamma^{e/o}_{\vec k}\equiv {1\over \sqrt{2WL}}\sum_{m,n}e^{-i(mk_x+2nk_y)}\gamma^{e/o}_{m,2n}\label{FT}\ee
with
\be \vec k=(2\pi r/L,\pi s/W),\label{k}\ee 
for integers $r$ and $s$. 
$m$ and $n$ run over $L$ and $W$ integer values, respectively, where $L$ and $2W$ are the length and width of the lattice in 
$x$ and $y$ directions, respectively. $k_x$ is
between $-\pi$ and $\pi$ while $k_y$ is between $-\pi /2$ and $\pi /2$. 
This gives
\be \{\gamma^{e/o}_{\vec k},\gamma^{e/o}_{\vec k'}\}=\delta_{\vec k,-\vec k'}.\ee
Inverting Eq. \eqref{FT}, we find
\be \gamma^{e/o}_{\vec r}=\sqrt{2\over WL}\sum_{\vec k}e^{i\vec k\cdot \vec r}\gamma^{e/o}_{\vec k},\ee
where $\vec r =(m,2n)$. 

The Hermiticity of $\gamma^{e/o}_{\vec r}$ implies that
\be \label{eq:herm}\gamma^{e/o}_{-\vec k}={\gamma^\dagger}^{e/o}_{\vec k}.\ee
We can then write the noninteracting Hamiltonian~\eqref{eq:H0} as 
\be
H_0=2it\sum_{\vec k}[(\gamma^e_{-\vec k}\gamma^e_{\vec k}-\gamma^o_{-\vec k}\gamma^o_{\vec k})e^{ik_x}+\gamma_{-\vec k}^e\gamma^o_{\vec k}(1-e^{-2ik_y})].
\ee
Using Eq.~\eqref{eq:herm}, we then obtain
\be
\begin{split} H_0=2it\sum_{k_x>0}[&(\gamma^{e\dagger}_{\vec k}\gamma^e_{\vec k}-\gamma^{o\dagger}_{\vec k}\gamma^o_{\vec k})(e^{ik_x}-e^{-ik_x})\\
&
+\gamma_{\vec k}^{e\dagger}\gamma^o_{\vec k}(1-e^{-2ik_y})-\gamma_{\vec k}^{o\dagger}\gamma^e_{\vec k}(1-e^{2ik_y})],
\end{split}
\label{H}\ee
where a constant was dropped.
Diagonalizing the noninteracting Hamiltonian~\eqref{H} gives the dispersion relation
\be E_\pm =\pm 4t\sqrt{\sin^2k_x+\sin^2k_y}.\label{Epm}\ee
There are Dirac points at $\vec k=(0,0)$ and $(\pi ,0)$ with two species of massless relativistic fermions  in their vicinity, i.e., $E_\pm=\pm 4t|\vec k|$, with the momentum measured from the Dirac points. Note that we must restrict to $0\leq k_x\leq \pi$ and 
$-\pi /2\leq k_y\leq \pi /2$. 

The second neighbor hopping adds a term:
\be  H_2=8it_2\sum_{k_x>0}\cos k_x\cos k_y[e^{-ik_y}\gamma^{e\dagger}_{\vec k}\gamma^o_{\vec k}
-e^{ik_y}\gamma^{o\dagger}_{\vec k}\gamma^e_{\vec k}],\ee
which changes the dispersion relation to
\be E_\pm =\pm \sqrt{(4t\sin k_x)^2+(4t\sin k_y)^2+(8t_2\cos k_x\cos k_y)^2}.\label{delta_H}\ee
Therefore, near the Dirac points, we obtain a massive relativistic dispersion relation:
\be E_\pm =\pm \sqrt{(4t)^2(k_x^2+k_y^2)+(8t_2)^2},\label{Epm_delta}\ee
where the momenta  are measured from the Dirac points.

\subsection{Topological classification}
To perform the analysis of the topological invariant of the Hamiltonian, Eqs. \eqref{H} and \eqref{delta_H}, we rewrite the total Hamiltonian in the matrix form:
\begin{align}
H = \begin{pmatrix}
\gamma^{e\dag}_k\\
\gamma^{o\dag}_k
\end{pmatrix}
\mathcal{H}
\begin{pmatrix}
\gamma^e_k & \gamma^o_k
\end{pmatrix}, 
\end{align}
where
\begin{align}
\mathcal{H} = d_x(k)\sigma^x+d_y(k)\sigma^y+d_z(k)\sigma^z. \label{eq:H}
\end{align}
Here $\sigma$'s are the usual Pauli matrices and  
\begin{align}
d_x(k) &= 4 t \sin k_y \cos k_y + 8 t_2 \cos k_x \cos k_y \sin k_y, \\
d_y(k) &= 4 t \sin^2 k_y - 8 t_2 \cos k_x \cos^2 k_y, \\
d_z(k) &= 4 t \sin k_x.
\end{align}

Phase transitions in the noninteracting model can be understood in terms of the closing of the spectral gap of $\mathcal{H}$. The energy \eqref{delta_H} can only vanish at $t_2=0$ and $k_x=0, \pi$, $k_y=0$ in the allowed range of the momenta. Thus there are two Majorana gap closings for $t_2=0$, which should correspond to a change of the topological invariant by $2$. 

We now proceed to computing the Chern number, $\mathcal{C}$, of Eq. \eqref{eq:H}, which is
\begin{align}
\mathcal{C} = \frac{1}{4\pi} \int d^2 k \frac{1}{|d(k)|^3} \mathbf{d}(k)\cdot \frac{\partial \mathbf{d}(k)}{\partial k_x}\times \frac{\partial \mathbf{d}(k)}{\partial k_y}.
\end{align}
Here the integration goes over the whole Brillouin zone. Explicit integration in this formula gives $+1$ for $t_2/t<0$, and $-1$ for $t_2/t>0$. The $t_2=0$ transition 
is thus topological with Chern number change of $\pm 2$. Such a phase transition hosts two Majorana cones, in accordance with the gap closing analysis above. We thus conclude that the $t_2=0$ point is a topological transition gapless point between $\mathcal{C}=\pm 1$ phases. 

The gapped phases for $t_2\neq 0$ are analogues of the $p\pm ip$ superconductors, or superconducting analogues of the quantum Hall phases with chiral Majorana (instead of complex fermion) modes at the edges. It is known that the Chern number classification survives in presence of interactions~[\onlinecite{Wen2004}], therefore the analysis of the interacting Hamiltonian with $t_2=0$ is equivalent to analyzing a topological transition between $\mathcal{C}=\pm 1$ topological superconducting phases.

 \section{Low Energy Effective Field Theory and its Emergent Symmetries{}}
 \label{FTS}
 \subsection{Low-energy Hamiltonian}
 We start with the $t_2=0$ noninteracting Hamiltonian $H_0$. Keeping only Fourier modes near the two gapless points, we write:
 \be \gamma^{e/o}_{\vec r}\approx 2\sqrt{2}[\chi^{e/o+}(\vec r)+(-1)^x\chi^{e/o-}(\vec r)],\label{chid}\ee
 where $\chi^{e/o\pm}(\vec r)$ vary slowly. [The $2\sqrt{2}$ factor in Eq. (\ref{chid}) is derived in Appendix B.]
 This gives the  anticommutators:
 \be \{\chi^i(\vec r),\chi^j(\vec r')\}={1\over 2}\delta^{ij}\delta^2(\vec r-\vec r'),\ee
 where $i$, $j$, label the four species of fermions, $\pm$, $e/o$. [This normalization of the anti-commutators is convenient because the relativistic Langrangian density 
  is consequently unit normalized.]
 
  Expanding to first order in derivatives, and using $\sum_{m,n}f(m,2n)\to {1\over 2}\int dxdyf(x,y)$, we find $H_0\approx\int dx dy{\cal H}_0$ with
 \be 
 \begin{split}{\cal H}_0 =4it\sum_{\pm}\{ \pm &[\chi^{e\pm}\partial_x\chi^{e\pm}-\chi^{o,\pm}\partial_x\chi^{o\pm}]\\
 &+\chi^{e\pm}\partial_y\chi^{o\pm}
 +\chi^{o\pm}\partial_y\chi^{e\pm}\}.
 \end{split}
 \ee
 
 Combining $\chi^{e\pm}$ and $\chi^{o\pm}$ into  2-component spinors, $\vec \chi^{+}\equiv (\chi^{e+},\chi^{o+})^T$, $\vec \chi^-\equiv (\chi^{o-},\chi^{e-})^T$, we can write
 \be {\cal H}_0=4it\sum_{\pm}\vec \chi^{T\pm} \cdot [\sigma^z\partial_x+\sigma^x\partial_y]\vec \chi^{\pm}.\ee

Going back to  momentum space, we can diagonalize the Hamiltonian in terms of the low energy fields. The Majorana nature of the fields implies  $\vec\chi_{-\vec k}^{\pm}=\vec\chi^{\pm \dagger}_{\vec k}$.  The Hamiltonian for the $\chi^+$ field, 
dropping the superscript, becomes
  \be H_0=v\int_{k_x>0} d^2k[\chi^{\dagger}_{\vec k,>}\chi_{\vec k,>}-\chi^{\dagger}_{\vec k,<}\chi_{\vec k,<}]|\vec k|,\ee
 with the velocity $v=4t$, and $\chi_{\vec k,>}$ ($\chi_{\vec k,<}$) representing the eigenmodes with positive (negative) energy. Now it is convenient to make a particle-hole transformation for the negative energy operators:
 \be \chi_{\vec k,<}\to \chi^\dagger_{-\vec k,<}.\ee
The sign change $\vec k\to -\vec k$ indicates that annihilating a fermion of momentum $\vec k$ changes the momentum by $-\vec k$. The Hamiltonian can then be written as 
 \be H_0=v\int_{k_x>0} d^2k[\chi^{\dagger}_{\vec k,>}\chi_{\vec k,>}+\chi^{\dagger}_{-\vec k,<}\chi_{-\vec k,<}]|\vec k|.\ee
As we are now covering the entire $\vec k$-space, with both signs of $k_x$, we may drop the subscripts and simply write:
 \be H=v\int  d^2k\chi^{\dagger}_{\vec k}\chi_{\vec k}|\vec k|,\ee
 which is the standard result for a Majorana fermion.  There are particle excitations for all values of $\vec k$ and no antiparticle states. The interaction term can also be readily written in terms of the low-energy degrees of freedom as
  \be {\cal H}_{int}=256g\chi^{e-}\chi^{e+}\chi^{o+}\chi^{o-}.\label{Hinc}\ee
 Finally, in the low-energy theory, the second-neighbor hopping term becomes:
 \be {\cal H}_2=16it_2[\chi^{e+}\chi^{o+}-\chi^{e-}\chi^{o-}].\label{mass0}
 \ee

%

 \subsection{Emergent Lorentz symmetry}
We now check that the above Hamiltonian is indeed Lorentz invariant. The Lagrangian density can be written as
 \be {\cal L}_{0}=i\sum_\pm{\vec \chi}^{T,\pm} \cdot \partial_t{\vec \chi}^\pm -{\cal H}_0= {\cal L}_{0,+}+{\cal L}_{0,-}.\ee
The Lagrangian densities ${\cal L}_{0,\pm}$ depend on the chiral fields ${\vec \chi}^\pm$ as
 \be {\cal L}_{0,\pm}=i\vec \chi^{T,\pm} \cdot \partial_t\vec \chi^\pm -4it\vec \chi^{T\pm} \cdot [\sigma^z\partial_x+\sigma^x\partial_y]\vec \chi^{\pm}\ee
 We define the Dirac $\gamma$-matrices
 \be \gamma^\mu\equiv (\sigma^y,-i\sigma^x,i\sigma^z),\ee
 such that the anticommutator $\{\gamma^\mu,\gamma^\nu\}=2\eta^{\mu \nu}\openone$, with $\eta^{\mu \nu}={\rm diag}(1,-1,-1)$ and $\openone$ is the $2\times 2$ identity matrix. We further define
\be\bar \chi \equiv \vec \chi^T\gamma^0\ee
 for each of the chiral fields. We also simplify the notation by using $\chi$ instead of  $\vec \chi$.
 Suppressing the $\pm$ superscript, the chiral Lagrangian densities take the form
 \be {\cal L}_0=i\bar \chi \gamma^\mu \partial_\mu \chi,\ee 
 where we have set the velocity, $v=4t$,  to one. 
 The above Lagrangian density is invariant under a Lorentz transformation:
 \be \chi \to e^{i\gamma^\mu a_\mu /2}\chi,\ee
 which is not unitary in general. In the special case where only $a_0\neq 0$, this becomes:
 \begin{widetext}
 \be
 \begin{split}\bar \chi \gamma^\mu \partial_\mu \chi \to \bar \chi e^{-i\sigma^ya_0/2}\gamma^\mu e^{i\sigma^ya_0/2}\partial_\mu \chi
 =i\bar \chi [\gamma^0\partial_t+\gamma^1(\cos a_0\partial_x-\sin a_0\partial_y)+\gamma^2(\cos a_0\partial_y+\sin a_0\partial_x)]\chi
 \end{split}
  \ee
  \end{widetext}
  corresponding to a spatial rotation, with $\chi$ having a nonzero spin. Similarly, $a_1$ and $a_2$ correspond to Lorentz boosts.

Similarly,  the interaction term \eqref{Hinc} can be written in an explicitly Lorentz invariant form 
 \be {\cal H}_{int}\propto (\bar \chi^+\chi^+)(\bar \chi^-\chi^-).\ee
The second neighbor hopping term becomes:
  \be {\cal H}_2=8t_2[\bar \chi^+\chi^+-\bar \chi^-\chi^-]\label{mass}
 ,\ee
 which is a Lorentz invariant mass term with $m^\pm =\pm 8t_2$ for the $\chi^\pm$ fields. 
 
A simple renormalization-group scaling argument indicates that the interactions are irrelevant in the low-energy theory. 
 The fermion fields have dimension 1 in (2+1) dimensions so the interaction term $H_{int}$ has dimension 4. The marginal dimension is 3, which makes the interactions irrelevant. As discussed in the next subsection, this phase also has an emergent $U(1)$ symmetry and a conserved charge. 
 Thus, the low-energy analysis above predicts an extended relativistic massless phase around the noninteracting point with $t_2=0$ in the presence of interactions. This phase extends to finite critical values of $g$, $g_{c3}<g<g_{c4}$, as summarized in Sec.~\ref{sec:intro}. [Actually, we find that the effective hopping strength in $x$ and $y$ directions become unequal for 
infinitesimal positive $g$. However, this does not eliminate the massless behavior and can be eliminated by a rescaling 
of the $y$ coordinate.]

  \subsection{Emergent $U(1)$ symmetry}

The $U(1)$ symmetry corresponds to a Dirac fermion obtained by combining the two Majorana modes as
 \be \psi \equiv \left(\begin{array}{c} \chi^{e+}+i\chi^{o-}\\ \chi^{o+}+i\chi^{e-}\end{array}\right).\label{Dirac}\ee
 This gives $\bar \psi \psi =-2i[\chi^{e+}\chi^{o+}-\chi^{e-}\chi^{o-}]$ and  
 \be (\bar \psi \psi )^2=8\chi^{e+}\chi^{o+}\chi^{e-}\chi^{o-}.\ee
 Setting $v=4t$ to 1, the Lagrangian density becomes:
 \be {\cal L}=\bar \psi (i\gamma^\mu \partial_\mu -m)\psi -32g(\bar \psi \psi )^2.
 \ee
There is an emergent $U(1)$ particle number conservation symmetry, $\psi\to e^{i\theta}\psi$, in addition to the emergent Lorentz symmetry.
While the irrelevant interaction term (of dimension 4) respects the Lorentz and $U(1)$ symmetries, we expect even higher dimension operators to be present in the effective Hamiltonian, 
which break the particle number symmetry, such as $\chi^{e+}\partial_x\chi^{e+}\chi^{o+}\partial_x\chi^{o+}$.

  Note that the emergent U(1) symmetry rotates
  \bea \left(\begin{array}{c} \chi^{e+}\\ \chi^{o-}\end{array}\right)&\to& R(\theta ) \left(\begin{array}{c} \chi^{e+}\\ \chi^{o-}\end{array}\right),\nonumber \\
\left(\begin{array}{c} \chi^{o+}\\ \chi^{e-}\end{array}\right)&\to& R(\theta ) \left(\begin{array}{c} \chi^{o+}\\ \chi^{e-}\end{array}\right),
  \eea
  where $R(\theta )$ is an SO(2) rotation matrix. A $\pi/2$ rotation in the field theory corresponds to an exact 
symmetry of the lattice model, translation  along a diagonal:
\be \gamma_{m,n}\to (-1)^m\gamma_{m+1,n+1},\label{Up}\ee
which corresponds to
\be \chi^{e+}\to -\chi^{o-},\ \  \chi^{o-}\to \chi^{e+},\ \  \chi ^{o+}\to -\chi^{e-},\ \  \chi^{e-}\to \chi^{0+}.\ee
Thus a subgroup of the $U(1)$ symmetry, consisting of rotations by $\pm \pi /2$, $\pi$ and $0$ is an exact 
symmetry of the lattice model. This symmetry, Eq. (\ref{Up}), remains with second-neighbor hopping present.  It is also respected by the Lorentz invariant 
interaction term.    

Note that this model avoids the fermion doubling problem~[\onlinecite{Nielson1981}]. We only have one Dirac fermion in the low-energy theory at the cost of $U(1)$ symmetry only being emergent instead of an exact lattice symmetry. Such Majorana models might be useful for high energy physics simulations [\onlinecite{O'Brien2017}].

\subsection{Other symmetries} 
We now consider what the other exact symmetries of the lattice model, discussed in Sec. \ref{symm}, correspond to the in the field theory. 

Translation by one site in the $x$ direction, an exact symmetry even in the presence of $t_2$, corresponds to 
\be \chi^{e/o\pm}\to \pm \chi^{e/o\pm}\ee
 and hence to the charge conjugation symmetry, C, with $\psi\to \psi^*$, which
takes the mass term into itself. 
Time reversal takes
  \bea \chi^{e+}&\leftrightarrow& \chi^{e-},\nonumber \\
  \chi^{o+}&\leftrightarrow&-\chi^{o-}.\eea
  or $\psi\to -\gamma^0\psi^*$. 
  This changes the sign of the mass term:
  \be \bar \psi \psi \to \psi^T\gamma^0\gamma^{0*}\gamma^0\psi^* =-\psi^T\gamma^0\psi^*=\psi^\dagger (\gamma^0)^T\psi
= -\bar \psi \psi ,
\ee
  as expected since it is violated by $t_2$. Parity symmetry interchanges
\be \chi^{e/o+}(x,y)\leftrightarrow \chi^{e/o-}(-x,y)\ee
corresponding to
\be \psi(x,y)\to -\gamma^1\psi^*(-x,y).\ee
In the field theory, this should be considered CP, a product of charge conjugation and parity: 
\be \psi (x,y)\to -\gamma^1\psi (-x,y).\ee
Parity (CP in the field theory) changes the sign of the mass term. 
So, we see that the model, including the mass term, is invariant under C but not P or T. It {\it is} invariant 
under PT (which flips the sign of the mass term twice) as expected from the CPT theorem.

  The spatial rotation by $\pi /2$ takes
  \bea \chi^{e+}+\chi^{e-}&\to& \chi^{e+}+\chi^{e-},\nonumber \\
  \chi^{e+}-\chi^{e-}&\to& \chi^{o+}+\chi^{o-},\nonumber \\
   \chi^{o+}+\chi^{o-}&\to&-(\chi^{e+}-\chi^{e-}),\nonumber \\
    \chi^{o+}-\chi^{o-}&\to& \chi^{o+}-\chi^{o-},
  \eea
  while rotating the spatial coordinates by $\pi /2$. 
  This gives 
  \bea \chi^{e+}&\to& {1\over 2}(\chi^{e+}+\chi^{e-}+\chi^{o+}+\chi^{o-}),\nonumber \\
  \chi^{e-}&\to& {1\over 2}(\chi^{e+}+\chi^{e-}-\chi^{o+}-\chi^{o-}),\nonumber \\
  \chi^{o+}&\to& {1\over 2}(-\chi^{e+}+\chi^{e-}+\chi^{o+}-\chi^{o-}),\nonumber \\
   \chi^{o-}&\to& {1\over 2}(-\chi^{e+}+\chi^{e-}-\chi^{o+}+\chi^{o-}).
  \eea
  On the other hand, in the field theory, a $\pi /2$ rotation just mixes the upper and lower components of the two independent Majorana spinors,
  \be  \chi \to e^{-i\pi \sigma^y/4}\chi,\ee
  as discussed above.  Thus
  \bea \left(\begin{array}{c} \chi^{e+}\\ \chi^{o+}\end{array}\right)&\to& {1\over \sqrt{2}}\left(\begin{array}{cc} 1&1\\ -1&1\end{array}\right)  \left(\begin{array}{c} \chi^{e+}\\ \chi^{o+}\end{array}\right),\nonumber \\
   \left(\begin{array}{c} \chi^{o-}\\ \chi^{e-}\end{array}\right)&\to& {1\over \sqrt{2}}\left(\begin{array}{cc} 1&1\\ -1&1\end{array}\right)  \left(\begin{array}{c} \chi^{o-}\\ \chi^{e-}\end{array}\right),\label{FTrot}
  \eea
  or 
  \bea \chi^{e+}&\to& {1\over \sqrt{2}}(\chi^{e+}+\chi^{o+}),\nonumber \\
  \chi^{e-}&\to& {1\over \sqrt{2}}(-\chi^{e-}+\chi^{o-}),\nonumber \\
  \chi^{o+}&\to& {1\over \sqrt{2}}(-\chi^{e+}+\chi^{o+}),\nonumber \\
  \chi^{o-}&\to& {1\over \sqrt{2}}(\chi^{e-}+\chi^{o-}).\nonumber \\
  \eea
  A $\pi /2$ spatial rotation in the lattice model corresponds to a $\pi/2$ spatial rotation in the field theory followed by the transformation
  \bea \left(\begin{array}{c} \chi^{e+}\\ \chi^{o-}\end{array}\right)&\to& {1\over \sqrt{2}}\left(\begin{array}{cc} 1&1\\ -1&1 \end{array}\right)
  \left(\begin{array}{c} \chi^{e+}\\ \chi^{o-}\end{array}\right),\nonumber \\
   \left(\begin{array}{c} \chi^{o+}\\ \chi^{e-}\end{array}\right)&\to& {1\over \sqrt{2}}\left(\begin{array}{cc} 1&1\\ -1&1 \end{array}\right)
  \left(\begin{array}{c} \chi^{o+}\\ \chi^{e-}\end{array}\right).\label{sttr}
  \eea
    This latter transformation is the $U(1)$ symmetry discussed above with rotation angle $-\pi /4$. So, we see that a spatial rotation by $\pi /2$ in the lattice 
    model corresponds to the product of a spatial rotation by $\pi /2$ and an internal symmetry rotation by $-\pi /4$: $\psi \to e^{-i\pi /4}e^{i\pi \gamma^0/4}\psi$.

\section{Mean-Field Treatment of the Phase Diagram of the Interacting Model}
\label{MFT_PD}
\subsection{Mean-field decoupling}
Similar to the 1D case, we expect this 2D model to have a rich phase diagram versus the one free parameter $g/t$.
In this section, we 
will predict a phase diagram using mean-field approximations to the lattice model. In the next section, we 
will analyze the various phases using the low-energy field theory and discuss the nature of the phase transitions. 
The interaction term can be factorized into horizontal, vertical, or diagonal nearest-neighbor factors:
\bea H_{int}&=&g\sum_{m,n}(i\gamma_{m,n}\gamma_{m+1,n})(i\gamma_{m,n+1}\gamma_{m+1,n+1})\nonumber \\
&=&-g\sum_{m,n}(i\gamma_{m,n}\gamma_{m,n+1})(i\gamma_{m+1,n}\gamma_{m+1,n+1})\nonumber \\
&=&-g\sum_{m,n}(i\gamma_{m,n}\gamma_{m+1,n+1})(i\gamma_{m,n+1}\gamma_{m+1,n})\label{Hintfac}
\eea
We thus consider three possible decouplings of the interaction term, where the expectation values are summarized for each case in the table below
\begin{center}
\begin{tabular}{|c|c|c|}
\hline 
Horizontal & Vertical & Diagonal\\ 
\hline 
$ \langle i\gamma_{m,n}\gamma_{m+1,n}\rangle$ &$ \langle i\gamma_{m,n}\gamma_{m,n+1}\rangle$ & $\langle i\gamma_{m,n}\gamma_{m+1,n\pm1}\rangle$ \\ 
\hline 
\end{tabular} 
\end {center}
As shown in Sec. \ref{symm},  the horizontal and vertical order parameters are rotated into each other by the $\pi /2$ spatial rotation symmetry.  They thus correspond to equivalent states, with this 
rotation symmetry spontaneously broken. Without loss of generality (due to the symmetry above), we restrict ourselves to vertical and diagonal decoupling, and do not explicitly study the horizontal decoupling case. We further assume that the unit cell is not larger than $2\times 2$. This implies that the order parameters have a general dependence on coordinates:
\be \label{eq:O}
O_{m,n}=A+B(-1)^n+C(-1)^m+D(-1)^{m+n}.
\ee

\subsection{$g>0$}
For $g>0$, the above factorizations in Eq. (\ref{Hintfac}) respectively suggest
\begin{eqnarray}
\langle i\gamma_{m,n+1}\gamma_{m+1,n+1}\rangle &=&-\langle i\gamma_{m,n}\gamma_{m+1,n}\rangle,\nonumber \\
 && (\text{antiferromagnetic horizontal order})\nonumber \\
\nonumber   \\
\langle i\gamma_{m+1,n}\gamma_{m+1,n+1}\rangle &=&\langle i\gamma_{m,n}\gamma_{m,n+1}\rangle \nonumber \\
&&  (\text{ferromagnetic vertical order})\label{eq:vert}\\ 
 \langle i\gamma_{m,n}\gamma_{m+1,n+1}\rangle &=&\langle i\gamma_{m,n+1}\gamma_{m+1,n}\rangle \nonumber \\
&& (\text{diagonal order}).
\label{eq:diag}
\end{eqnarray}

\subsubsection{Vertical Order}
\begin{figure}
\centerline{\includegraphics[width=\columnwidth]{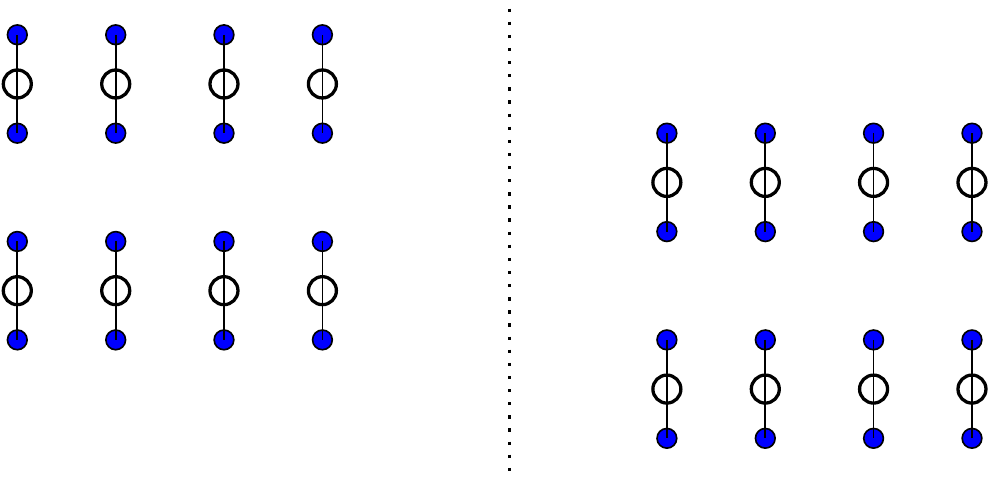}}
\caption{Sketch of the two mean-field ground states occurring for $g > 0$ and {$t> 0$}. The blue dots represent the lattice sites. The open circles appear on bonds on which the two Majorana modes combine to form Dirac fermions, which are unoccupied (lowering the energy for nanvanishing $t$). In addition, there are two equivalent states with Dirac fermions occurring on horizontal bonds. }
\label{VF}
\end{figure}

We first consider the case of vertical order. Equation~\eqref{eq:vert} then requires $C=D=0$, and the general order parameter can be written as
\be \bra i\gamma_{m,n}\gamma_{m,n+1}\ket=\langle i\gamma_{m+1,n}\gamma_{m+1,n+1}\rangle=A+B(-1)^n.\label{eq:vert2}\ee
{ Assuming $t>0$, we see from Eq. \eqref{elD} that $A<0$ term reduces the energy.} This term corresponds to enhanced vertical hopping. The $B$ term breaks not only rotational symmetry but also translation symmetry (in the vertical direction).  For $B<0$, 
pairs of Majorana fermions at sites $(m,2n)$ and $(m,2n+1)$ couple more strongly together than other nearest-neighbor pairs, {while for $B>0$, we have degenerate states with Majorana fermions at sites $(m,2n-1)$ and $(m,2n)$ coupling more strongly. The energy is insensitive to the sign of $B$, but $A$ must have the opposite sign to the sign of $t$ to minimize the energy. In the special case of $t=0$, the two signs of $A$ become degenerate and the degeneracy is doubled, with both empty and occupied Dirac fermions allowed in Fig. 2. } We may think of this as a tendency for pairs of 
nearest-neighbor Majoranas to pair into ordinary complex, i.e., Dirac, fermions, 
\be c_{m,n}\equiv (\gamma_{m,2n}+i\gamma_{m,2n+1})/2,\ee
with 
\be \bra i\gamma_{m,2n}\gamma_{m,2n+1}\ket=\bra(2c^\dagger_{m,n}c_{m,n}-1)\ket<0,\ee
corresponding to the resulting energy level tending to be empty.
This is illustrated in Fig. (\ref{VF}). 
Alternatively, this pairing could occur on sites $(m,2n-1)$ and $(m,2n)$. 
This is reminiscent of the broken translational symmetry that was found to occur in the 1D model~[\onlinecite{Rahmani2015, Rahmani2015a}]. This phase 
can be seen to preserve translation in the horizontal direction, time reversal and parity. 

As for the diagonal decoupling~\eqref{eq:diag}, the two terms have a different structure, and all parameters $A,B,C$, and $D$ could be potentially nonzero. All these parameters correspond to the uniform and staggered parts of interaction-induced diagonal hopping.

To compare these candidate ground states and estimate the phase diagram, we introduce order parameters into the Hamiltonian. 
For the case of the vertical order parameter, we first rewrite $H_{int}$ as:
\bea H_{int}&=&-g\sum_{m,n}[i\gamma_{m,n}\gamma_{m,n+1}-A -B (-1)^n]\nonumber \\
&\cdot& [i\gamma_{m+1,n}\gamma_{m+1,n+1}-A -B (-1)^n]\nonumber \\
&& -2g\sum_{m,n}[A +B (-1)^n]i\gamma_{m,n}\gamma_{m,n+1}\nonumber \\
&+&g2WL[A^2+B^2].\eea
Notice that the alternating term $2AB(-1)^n$ in $[A+B(-1)^n]^2$ vanishes upon summation over $n$ and does not enter the expression above.
 We then make the mean-field approximation of ignoring 
fluctuations, dropping the first term. The parameters $A$ and $B$ are then determined by minimizing the ground state 
energy of the resulting noninteracting Hamiltonian. 
Separating the constant proportional to $g(A^2+B^2)$, the rest of the Hamiltonian can be easily diagonalized by modifying the $\gamma^{e\dagger}_{\vec k}\gamma^o_{\vec k}$ term in Eq. (\ref{H}) and we obtain
\be H^v=2i\sum_{k_x>0}\gamma^{e\dagger}_{\vec k}\gamma^o_{\vec k}[(t-2gA-2gB )-(t-2gA
+2gB)e^{-2ik_y}].\ee
The energy levels of Eq. (\ref{Epm}) are thus modified to:
\be E_\pm \to \pm \sqrt{(4t\sin k_x)^2+[(4t-8gA )\sin k_y]^2+(8gB \cos k_y)^2}.\label{elD}\ee
Using $\sum_{\vec k}\to 2LW\int {d^2k\over (2\pi )^2}$,
the ground-state energy density then becomes
\be {E_0\over 2WL}\to g(A^2+B^2)+{1\over (2\pi )^2}\int_0^\pi dk_x\int_{-\pi /2}^{\pi /2}dk_y 
E_-(\vec k),\label{EDel}\ee
with $E_-(\vec k)$ defined in Eq. \eqref{elD}. Since there is a term linear in $A$ inside the square root in $E_-(\vec k)$, $A$ becomes nonzero for any $g>0$. To first order in $g$,
setting $B =0$,  we obtain the following ground-state energy per unit area (dropping a constant term):
\be {\cal E} \equiv {E_0\over 2WL}\approx 
gA^2+{2gA \over \pi^2}\int d^2k{\sin^2k_y\over \sqrt{\sin^2k_x+\sin^2k_y}}.\label{eq:vert_pos_en}\ee
Minimizing the energy density $\cal E$, we see that, as $g\to 0^+$,
\be A \to -{1\over \pi^2}\int d^2k{\sin^2k_y\over \sqrt{\sin^2k_x+\sin^2k_y}}.\ee
There is a first-order phase transition into the phase with broken rotational symmetry, but unbroken 
translational symmetry, at $g=0$.

\begin{figure}
\centerline{\includegraphics[width=\columnwidth]{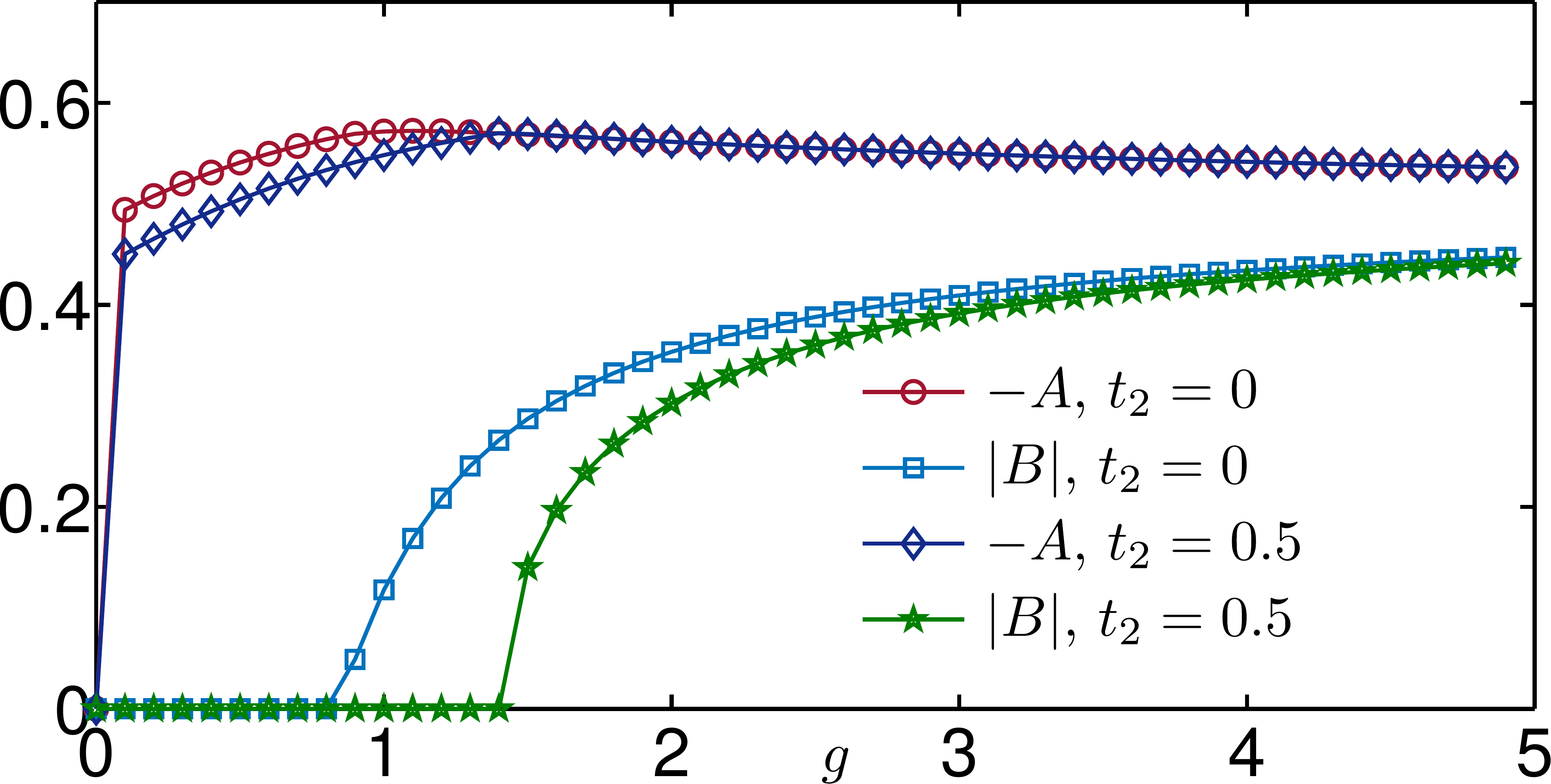}}
\caption{The values of $A$ and $B$ that minimize the energy density~\eqref{EDel} ($t_2=0$), as a function of $g$. In addition to a first-order transition at $g=0^+$, we find a critical point near $g\approx 0.9$. For a finite $t_2=0.5$, we also find a critical point and a phase transition to an emergent superfluid phase $|B|>0$. However, this transition is in a different universality class than the $t_2=0$ case as the normal phase is gapped (gapless) for $t_2\neq 0$ ($t_2=0$).}
\label{f2}
\end{figure}

We then numerically minimize the above energy density over both $A$ and $B$. The results are shown in Fig. (\ref{f2}). The $B$ order parameter remains zero until a second-order phase transition, where it becomes nonzero, at $g\approx 0.9$ for $t_2=0$. The presence of diagonal hopping $t_2\neq 0$ adds another term, $(8t_2\cos k_x \cos k_y)^2$ [see Eq.~\eqref{delta_H}], under the square root in the dispersion relation $E_\pm$. Again, the energy can be directly minimized over $A$ and $B$.  We find a very similar behavior, as shown, e.g., for $t_2=0.5$ in Fig.~(\ref{f2}). Indeed, there is a critical point at which $B$ becomes nonzero. As 
shown in Sec. (\ref{sec:ft}), in the large $g$ phase the emergent $U(1)$ symmetry is spontaneously broken.  While this
symmetry breaking transition occurs for both $t_2$ zero and nonzero, the universality class of the phase transition is 
quite different in the two cases, as also discussed in  Sec. (\ref{sec:ft}).

To further analyze the second-order phase transition above for $t_2=0$, we set $B=0$ and find $A_c(g)$, the value of $A$, which minimizes the energy when $B=0$, as a function of $g$. 
To determine when a vanishing $B$ stops minimizing the energy, we then compute the derivative of the ground-state energy 
per unit length, 
${\cal F}(g)\equiv{\partial\over \partial B^2}{\cal E}(A_c(g),B)
\Big |_{B=0}$ as follows: 
\begin{equation}
{\cal F}(g)=g-{2g^2\over \pi^2}\int d^2 k{\cos^2k_y\over\sqrt{[1 -2gA_c(g)]^2\sin^2 k_y+\sin^2 k_x}}\label{Fg}
\end{equation}
\begin{figure}[]
	\includegraphics[width=\columnwidth]{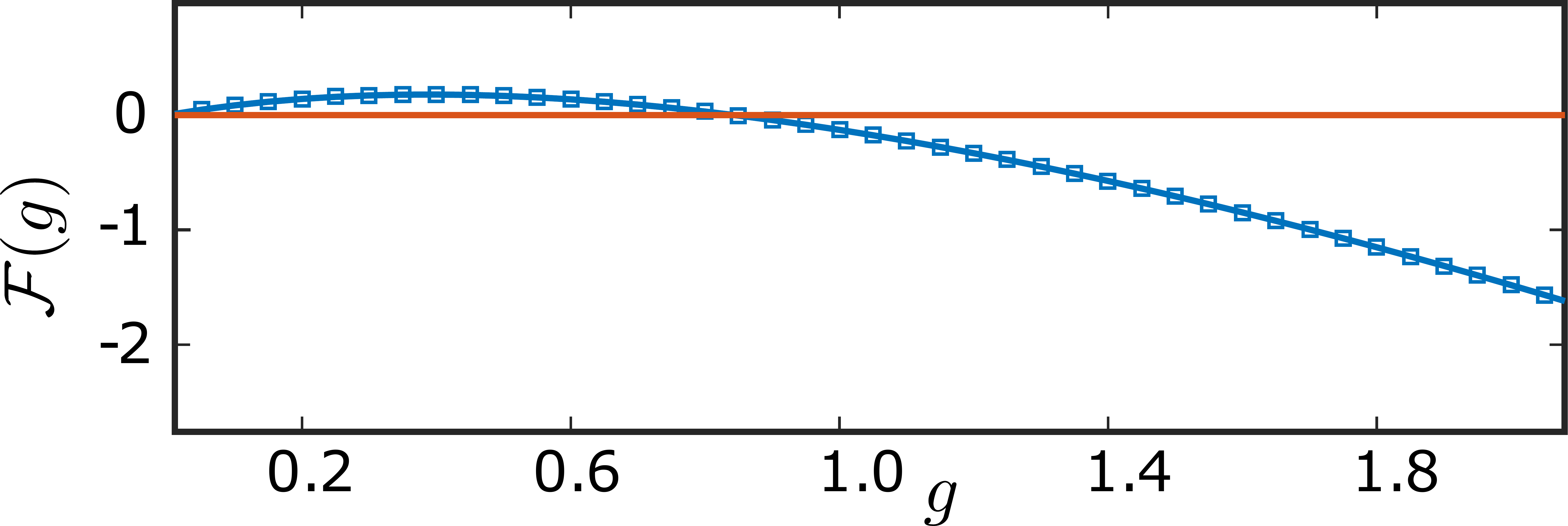}
	\caption{${\cal F}(g)$, Eq.~\eqref{Fg}, as a function of $g$. The critical point corresponds to ${\cal F}(g)=0$.}
	\label{f3}
\end{figure}
The dependence of ${\cal F}(g)$ on $g$ is shown in Fig. (\ref{f3}). The critical value of $g$, for which $B=0$ is no longer a mean-field solution is given by ${\cal F}(g_c)=0$, where this derivative changes sign from positive to negative. We find
\begin{equation}
g_c\approx 0.852851, \quad A_c(g_c)=-0.566753.
\end{equation}
We can further compute ${\cal F}(A_c(g_c),B)$ at $g=g_c$ for small $B$. The results are shown in Fig. (\ref{fgsmall}), indicating a linear dependence on $B$ for small $B$. As discussed in Sec.~\ref{sec:ft}, this linear term is determined by 
the low energy physics and can be obtained from the field theory approximation.

Finally, if $t_2$ is added to the gapless ($B=0$) mean-field Hamiltonian, we have checked that the Chern number remains the same ($\mathcal{C} = - \mathrm{sgn}\;(t_2/t)$), given that $A<t/2g$. Therefore for $0<g<g_{c4}$ the gapless line remains the transition between Chern number $1$ and $-1$ states in the mean-field description. Nonzero $B$ gives a gapped phase at zero $t_2$, with Chern number $0$.
\begin{figure}[]
	\includegraphics[width=\columnwidth]{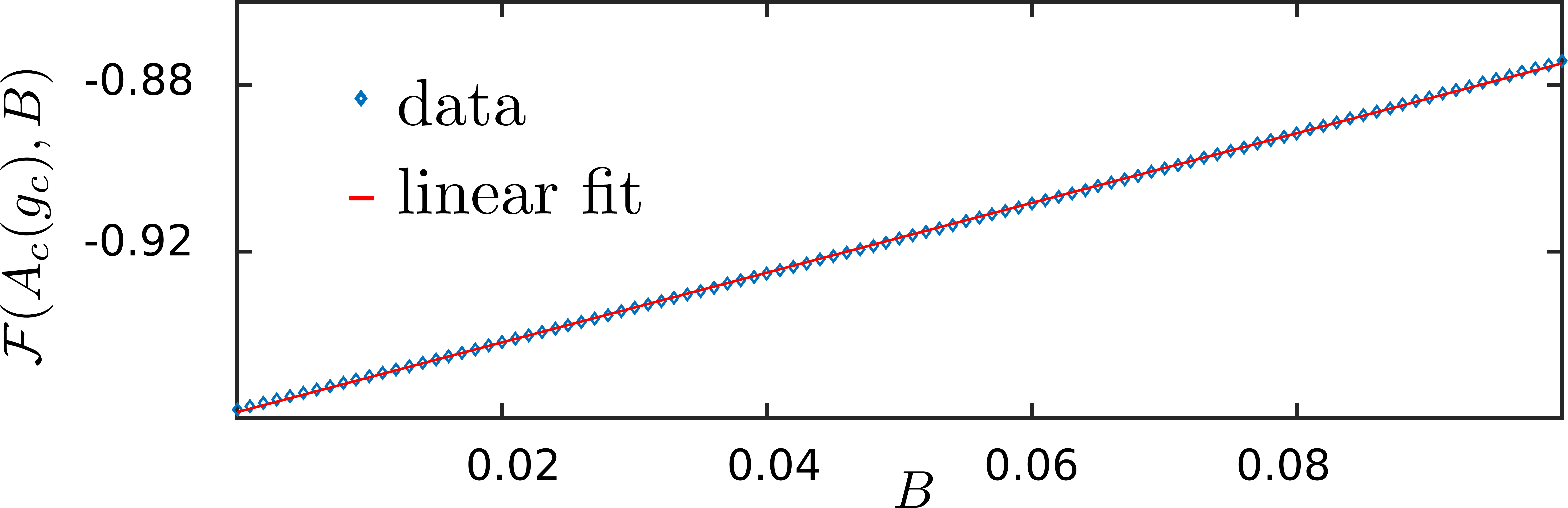}
	\caption{${\cal F}(A_c(g_c),B)$ for small $B$.}
	\label{fgsmall}
\end{figure}

\subsubsection{Diagonal order}
\begin{figure}[]
	\includegraphics[width=\columnwidth]{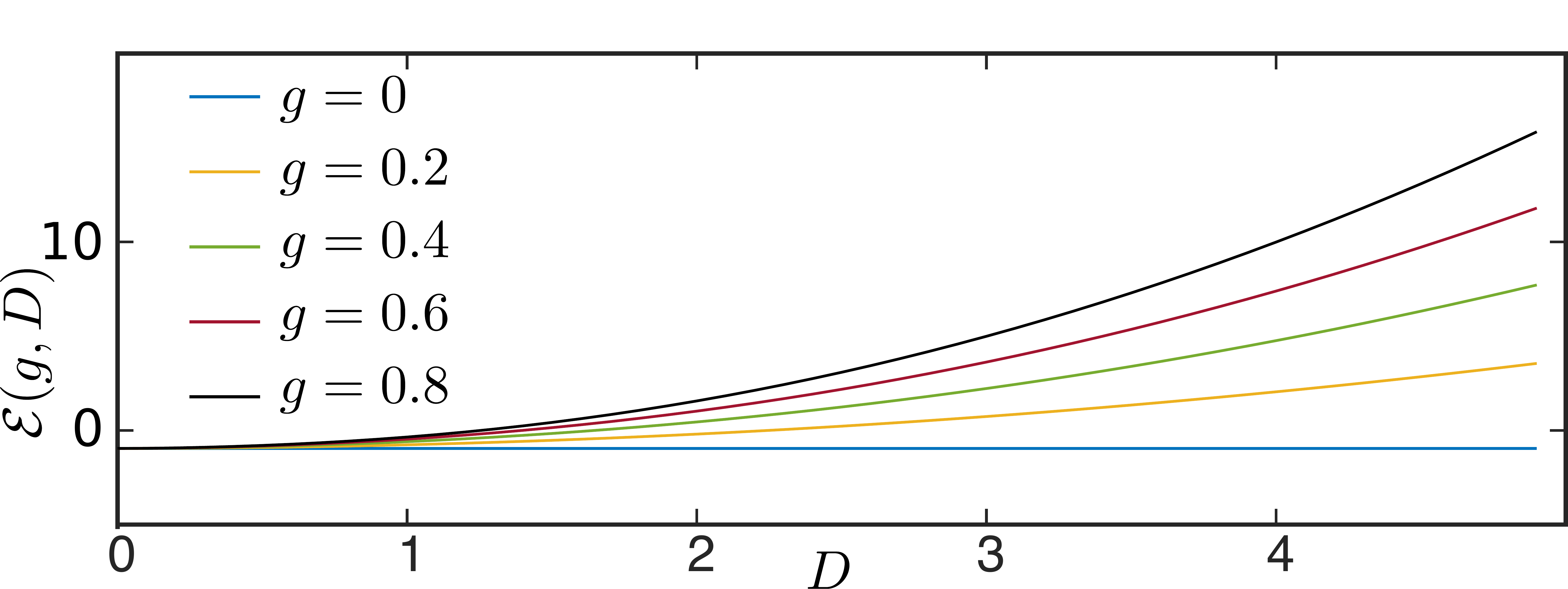}
	\caption{The energy density~\eqref{eq:ed} for $A=B=C=0$ as a function of $D$ for various values of $g$. The minimum energy always corresponds to $D=0$.}
	\label{f1}
\end{figure}

We now consider the diagonal order. We follow the same mean-field procedure, rewriting $H_{int}$ in terms of the 4 order parameters 
introduced in Eq.~\eqref{eq:O} and ignoring fluctuations, as above. The resulting addition to the noninteracting Hamiltonian is:
\begin{eqnarray}
H^d(A,B,C,D)&=&g\sum_{m,n}[-i O_{m,n}(\gamma_{m,n+1}\gamma_{m+1,n}\nonumber 
\\
&+&\gamma_{m,n}\gamma_{m+1,n+1})+O^2_{m,n}].
\end{eqnarray}
The constant term now gives 
\be
g\sum_{m,n}O_{m,n}^2=2WLg(A^2+B^2+C^2+D^2), 
\ee
as all terms with an alternating factor vanish upon summation. The rest of the Hamiltonian can be similarly diagonalized although the dispersion relation is lengthy and not very illuminating. Computing the total energy density and minimizing it over the four variables $A,B,C$, and $D$, we find numerically that the minimum energy occurs at $A=B=C=D=0$ for positive $g$. As an example, we plot the energy density as a function of $D$ for $A=B=C=0$ for various values of $g>0$, and observe that a nonzero $D$ only increases the energy, as shown in Fig.~(\ref{f1}).

\begin{figure}
\centerline{\includegraphics[width=0.5\columnwidth]{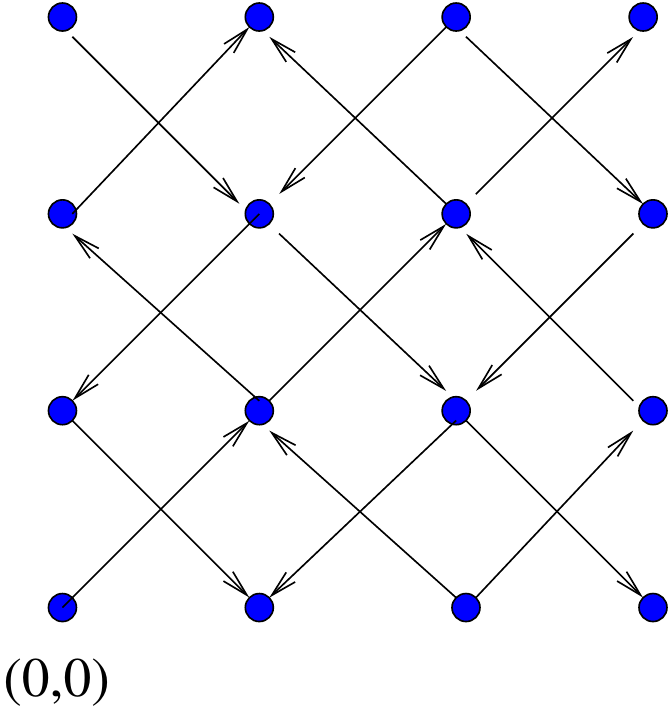}}
\caption{Sketch of the mean-field ground state arising from diagonal factorization of the interaction term for $g>0$, 
for only $D\neq 0$. An 
arrow pointing along a diagonal from site $\vec r_1$ to site $\vec r_2$ indicates that 
$i\langle \gamma_{\vec r_1}\gamma_{\vec r_2}\rangle >0$.}
\label{DgP}
\end{figure}

Indeed, from the low-energy field theory (see Sec. VI for details), we expect our mean-field ansatz to potentially only contain a nonvanishing $D$ term in this case:
\be \bra i\gamma_{m,n}\gamma_{m+1,n+1}\ket=\bra i\gamma_{m,n+1}\gamma_{m+1,n}\ket=D(-1)^{m+n},\label{DG0}\ee
We sketch this state in Fig. (\ref{DgP}), for one sign of $D$. This state breaks translation 
symmetry in the $x$ direction and parity symmetry while preserving translation symmetry in the 
$y$ direction, time reversal, and $\pi /2$ rotation symmetry.  It {\it does not} have an interpretation in terms of 
Majorana fermions pairing to form Dirac fermions. We can show analytically that for $g>0$, $D=0$ at the minimum energy and the vertical decoupling, which is symmetry related to horizontal decoupling, is energetically favorable. 

As before, we obtain the mean-field Hamiltonian by inserting Eq.~\eqref{DG0} into the expression for $H^d_{int}$. In momentum space, the Hamiltonian becomes
\be H^d_{int}\approx 8igD\sum_{\vec k}\gamma^e_{-\vec k+\pi \hat x}\gamma^o_{\vec k}e^{-ik_y}\cos k_x \cos k_y+2LWg{D^2}
.\ee
It is now convenient to make the replacement $\gamma_{\vec k}\to \gamma_{-\vec k}^\dagger$
for $k_y<0$. The Hamiltonian then becomes
\begin{widetext}
\be H=2WLg{D^2}+\sum_{k_x,k_y>0}\left(\gamma^{e\dagger}_{\vec k},\gamma^{e\dagger}_{\vec k-\pi \hat x},
\gamma^{o\dagger}_{\vec k},\gamma^{o\dagger}_{\vec k-\pi \hat x}\right){\mathscr H}
\left(\begin{array}{c}\gamma^e_{\vec k}\\ \gamma^e_{\vec k-\pi \hat x}\\ \gamma^o_{\vec k}\\ 
\gamma^o_{\vec k-\pi \hat x}\end{array}\right) .
\ee
where the matrix $\mathscr H$ is
\be {\mathscr H}=\left(\begin{array}{cccc} -4t\sin k_x&0&-4te^{-ik_y}\sin k_y&-8igD e^{-ik_y}\cos k_x\cos k_y\\
0& 4t\sin k_x&8igD e^{-ik_y}\cos k_x\cos k_y&-4te^{-ik_y}\sin k_y\\
-4te^{ik_y}\sin k_y&-8igD e^{ik_y}\cos k_x\cos k_y&-4t\sin k_x&0\\
8igD e^{ik_y}\cos k_x\cos k_y&-4te^{ik_y}\sin k_y&0&4t\sin k_x
\end{array}\right),\ee
with the four eigenvalues
\begin{eqnarray} E^\pm_1&=& \pm 4t\sqrt{\sin^2k_x+\sin^2k_y}+8gD \cos k_x\cos k_y,\nonumber \\
 E^\pm_2&=& \pm 4t\sqrt{\sin^2k_x+\sin^2k_y}-8gD \cos k_x\cos k_y.\label{Eposdiag}
 \end{eqnarray}
This gives the ground-state energy density:
\begin{equation}\label{eq:ed}
{\cal E}(g,D)={gD^2}-{t\over \pi ^2}\int_0^\pi d k_x \int_{0}^{\pi/2} d k_y\left[|\sqrt{\sin^2 k_x+\sin^2 k_y}+ 2 {g\over t}D\cos k_x \cos k_y|+|\sqrt{\sin^2 k_x+\sin^2 k_y}- 2  {g\over t}D\cos k_x \cos k_y|\right].
\end{equation}
\end{widetext}

Dropping a constant, the integral can be seen to be cubic (linear) in $|D |$ for small (large) $|D|$. This can be seen by noting that $
|a+b|+|a-b|=2|a|$ for $|a|>|b|$. Therefore, for large $D$, the term in the integral dominates and we get a linear dependence. For small $D$, we can get a $D$-dependent contribution only from a small region in momentum space $\sqrt{\sin^2 k_x+\sin^2 k_y}\approx |\vec k|<2gD/t$, leading to a cubic dependence on $D$. Therefore, unsurprisingly, 
as shown in Fig. (\ref{f1}), the minimum always occurs at $D=0$.
Thus the first mean-field state, with order parameter given in Eq. (\ref{eq:vert2}), always has lower energy for $g>0$.

\subsection{$g<0$}
For negative $g$, $H_{int}$ as written in Eq. (\ref{Hintfac}) suggests the following order parameters:
\begin{eqnarray}
\langle i\gamma_{m,n+1}\gamma_{m+1,n+1}\rangle=\langle i\gamma_{m,n}\gamma_{m+1,n}\rangle,\nonumber \\
\langle i\gamma_{m+1,n}\gamma_{m+1,n+1}\rangle=-\langle i\gamma_{m,n}\gamma_{m,n+1}\rangle,\label{eq:vert-}\\
 \langle i\gamma_{m,n}\gamma_{m+1,n+1}\rangle=-\langle i\gamma_{m,n+1}\gamma_{m+1,n}\rangle.\label{eq:diag-}
\end{eqnarray}
Again,  vertical and horizontal decouplings are related by $\pi /2$ spatial rotation symmetry so we just consider the case of 
vertical order, and compare it with the diagonal decoupling.

\subsubsection{Vertical order}
\begin{figure}
\centerline{\includegraphics[width=\columnwidth]{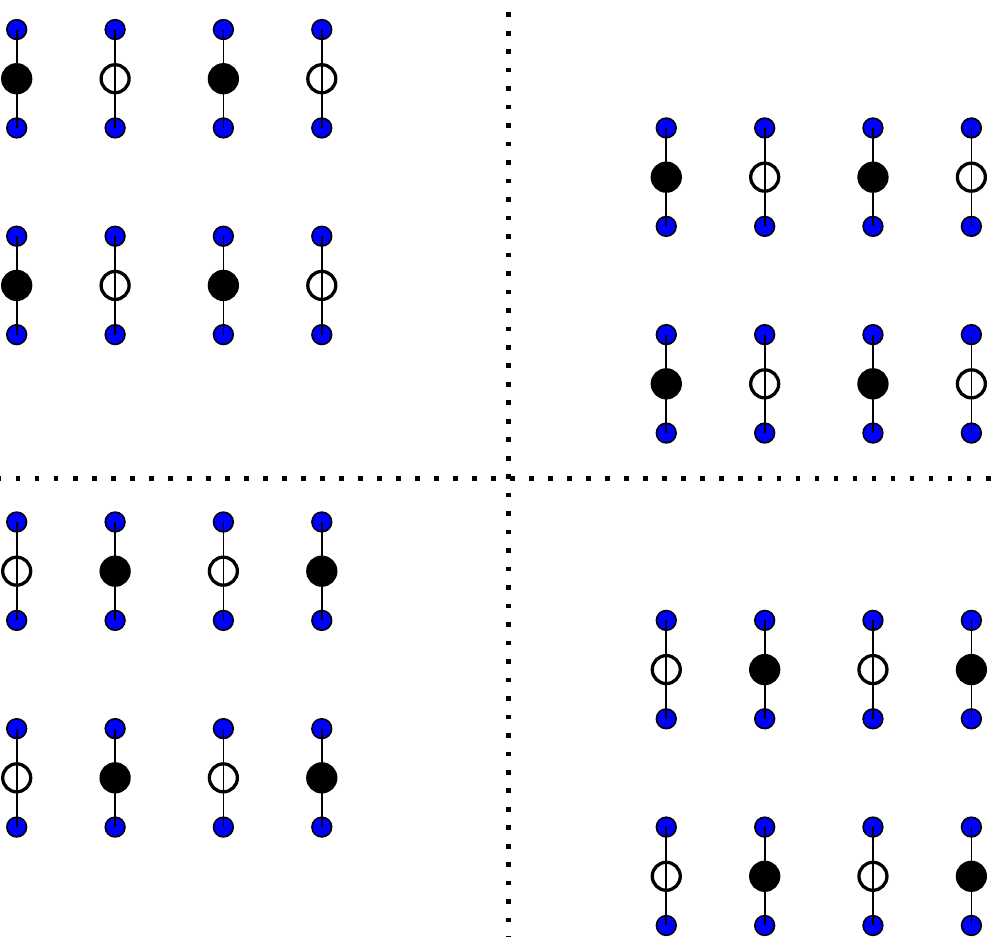}}
\caption{Sketch of four mean field ground states occurring for $g < 0$. The blue dots
represent lattice sites. The large circles appear on bonds on which the
two Majorana modes combine to form Dirac fermions, with filled or empty
circles corresponding to the Dirac level being occupied or empty. In addition, there are four equivalent ground states with Dirac fermions
occurring on horizontal bonds.}
\label{AVF}
\end{figure}

For a $2\times 2$ unit cell, Eq.~\eqref{eq:vert-} now corresponds to
\be \bra i\gamma_{m,n}\gamma_{m,n+1}\ket =C(-1)^m+D(-1)^{m+n},\label{VAF}\ee
which breaks translational symmetry (in both $x$ and $y$ directions) as well as rotational symmetry, while preserving parity and time reversal. 
The $D$ term preserves only translational symmetry on diagonals: $\gamma_{m,n}\to \gamma_{m+1,n+1}$ while the $C$ 
term preserves translation symmetry in the $y$-direction.  The $C$ term simply increases or decreases the vertical hopping on alternating columns, while the $D$ term can create a dimerization pattern with stronger bonds forming Dirac fermions as illustrated in Fig.~(\ref{AVF}). This is analogous to the effect of the $B$ term in the $g>0$ case.

\begin{figure}[]
	\includegraphics[width=\columnwidth]{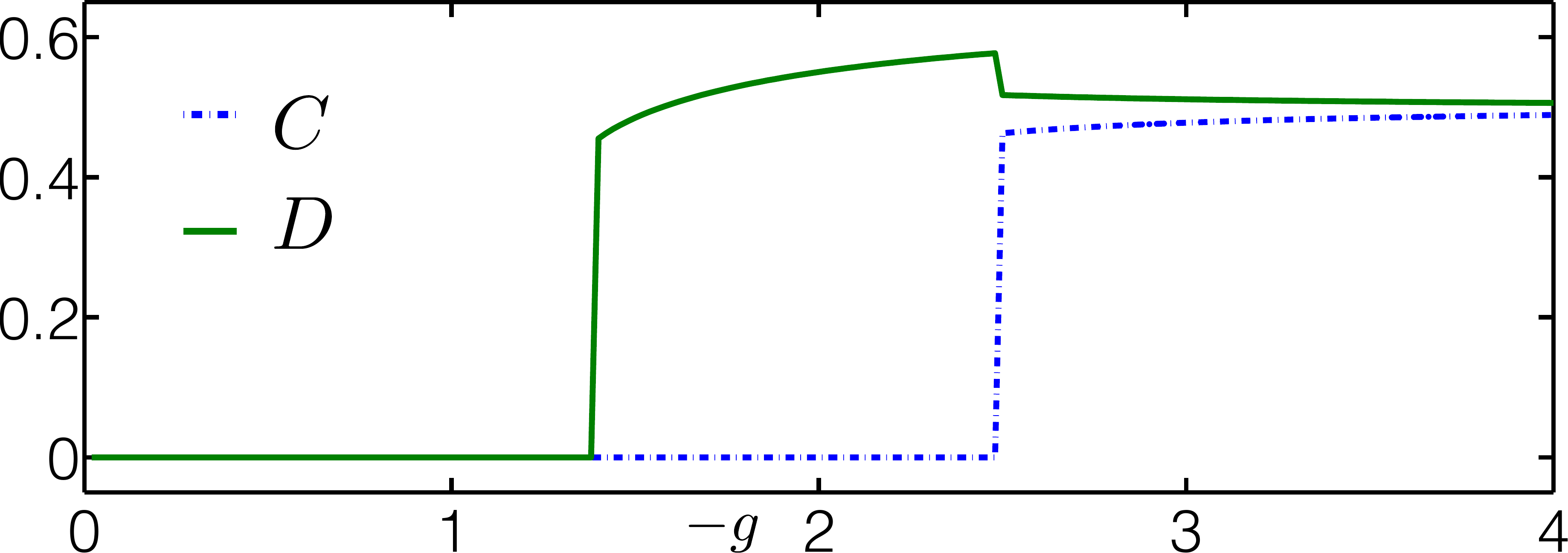}
	\caption{The mean-field parameters minimizing the energy density for the vertical decoupling case with $g<0$. There appears to be two mean-filed transitions; first to a solution with only nonvanishing $D$ and then to a state with both $D$ and $C$ nonzero.}
	\label{ss}
\end{figure}

\begin{figure}[]
	\includegraphics[width=\columnwidth]{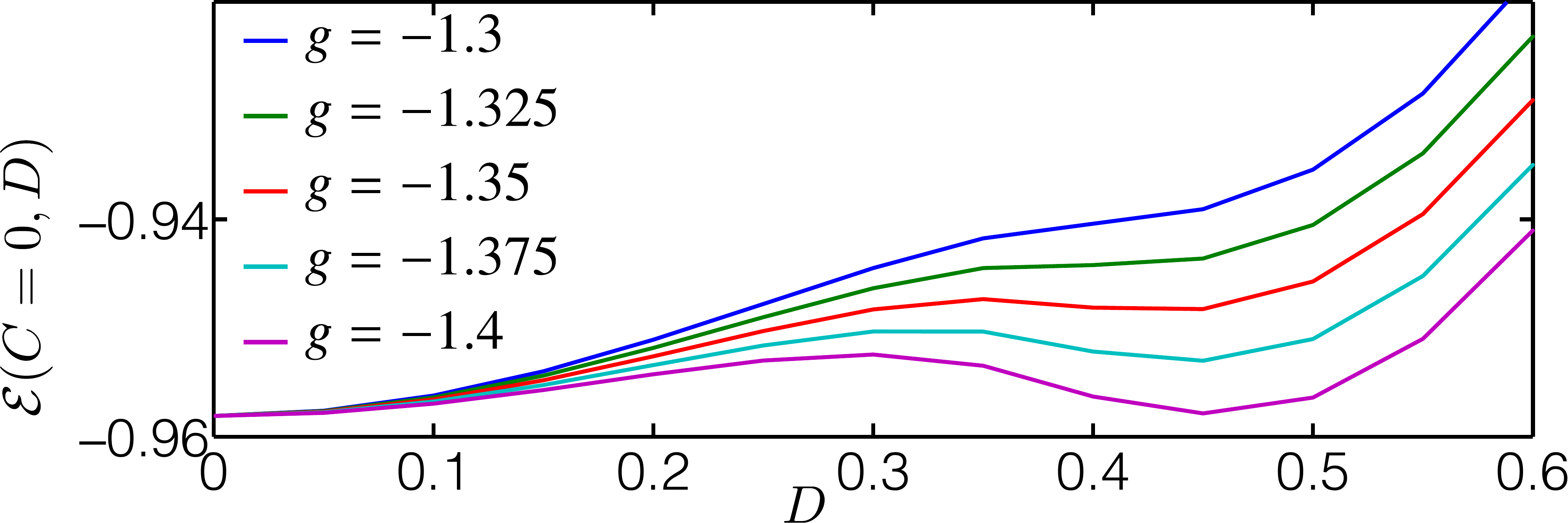}
	\caption{The energy density for various values of $g$ and $C=0$ as a function of $D$, supporting a first-order mean-filed transition.}
	\label{fg}
\end{figure}

The mean-field interaction Hamiltonian corresponding to Eq.~\eqref{VAF} can be written as
\be \begin{split}H^v_{int}= 2gi\sum_{m,n}&(-1)^m\gamma^e_{m,2n}[(D+C)\gamma^o_{m,2n}+(D-C)\gamma^o_{m,2n-2}]\\
&-2LWd(C^2+D^2).\end{split}\ee
\begin{widetext}
Upon Fourier transformation, the Hamiltonian becomes
\be H^v=-2WLg(C^2+D^2)+\sum_{k_x,k_y>0}\left(\gamma^{e\dagger}_{\vec k},\gamma^{e\dagger}_{\vec k-\pi \hat x},
\gamma^{o\dagger}_{\vec k},\gamma^{o\dagger}_{\vec k-\pi \hat x}\right){\mathscr H}
\left(\begin{array}{c}\gamma^e_{\vec k}\\ \gamma^e_{\vec k-\pi \hat x}\\ \gamma^o_{\vec k}\\ 
\gamma^o_{\vec k-\pi \hat x}\end{array}\right),
\ee
where the matrix $\mathscr H$ is now
\be \mathscr H=8g\left(\begin{array}{cccc} -{t\over 2g}\sin k_x&0&-{t\over 2g}e^{-ik_y}\sin k_y&e^{-ik_y}[iD\cos k_y+C\sin k_y]\\
0& {t\over 2g}\sin k_x&e^{-ik_y}[iA\cos k_y+B\sin k_y]&-{t\over 2g}e^{-ik_y}\sin k_y\\
-{t\over 2g}e^{ik_y}\sin k_y& e^{ik_y}[-iD\cos k_y+C\sin k_y]&-{t\over 2g}\sin k_x&0\\
 e^{ik_y}[-iD\cos k_y+C\sin k_y]&-{t\over 2g}e^{ik_y}\sin k_y&0&{t\over 2g}\sin k_x
\end{array}\right).\ee
This gives energy eigenvalues $\pm E_{1,2}$ with
\be E_{1,2}=4\sqrt{(t\sin k_x)^2+(t\sin k_y)^2+(2Dg\cos k_y)^2+(2Cg\sin k_y)^2\pm 4t\sqrt{(\sin k_x)^2(D\cos k_y)^2+(\sin^2k_x+\sin^2k_y)(C\sin k_y)^2}}.
\ee

\end{widetext}
The ground-state energy density (in the thermodynamic limit) is then given by
\be {\cal E}=-g(C^2+D^2)-{1\over (2\pi )^2}\int_0^\pi dk_x\int_0^{\pi /2}dk_y\sum_{i=1}^2E_i(\vec k).\ee
Minimizing the energy density above as a function of $C$ and $D$ gives two first-order phase transitions, when considering the vertical decoupling (we need to compare the energy of these phases with states having diagonal order). Upon increasing $-g$, we first have a phase transition to a state with nonvanishing $D$, which breaks the translation symmetry in the $y$ direction. Diagonal translation is still a symmetry in this phase. Increasing $-g$ further causes a phase transition to a phase with both $C$ and $D$ nonzero, which breaks the translation symmetry in the diagonal direction. In addition to the sudden jump in the order parameters [seen in Fig.~(\ref{ss})], the first-order nature of the phase transitions is confirmed by studying how the energy minima appear. As an example, we plot in Fig.~(\ref{fg}), the energy density as a function of $D$ for $C=0$  for several values of $g$. The energy of a  local minimum at finite $D$ decreases upon increasing $-g$ and crosses the energy for $D=C=0$ at the first-order phase transition. These phases were obtained by assuming a vertical mean-field decoupling. They may, however, be less favorable than the phases obtained by assuming diagonal decoupling. We will compare the energies after studying the diagonal decoupling case.

%
%
%
%

\subsubsection{Diagonal Order}

\begin{figure}[]
	\includegraphics[width=\columnwidth]{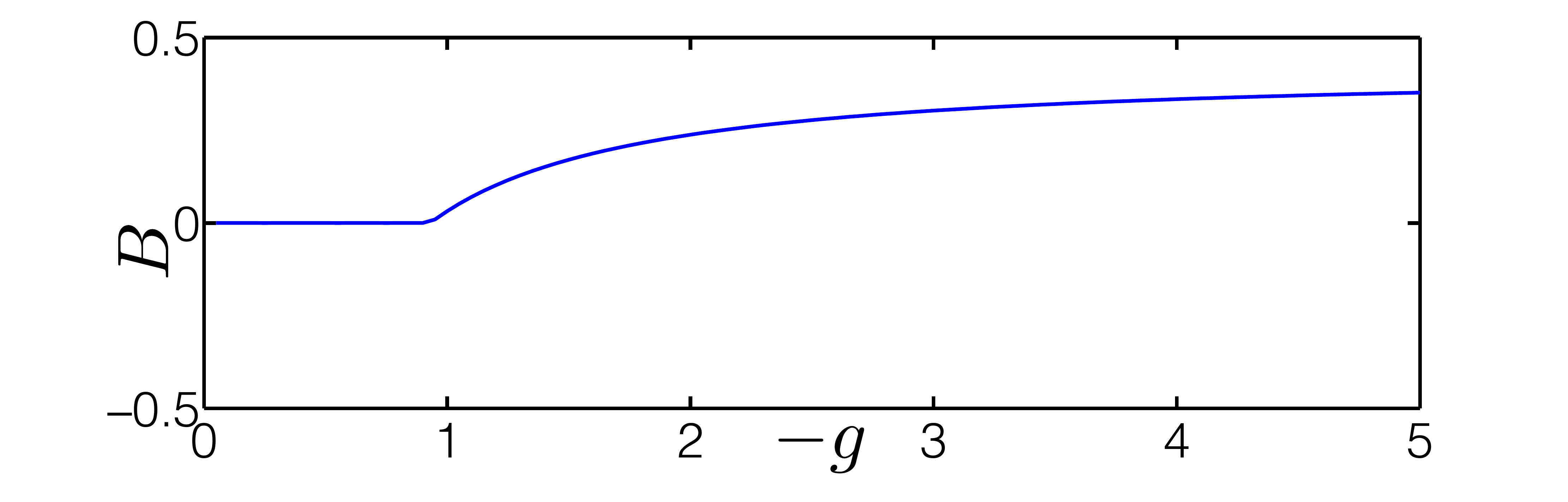}
	\caption{The value of $B$ that minimizes the energy density for the case of diagonal decoupling with $g<0$, indicating a second-order mean-field transition. The minimization yields $A=C=D=0$ for all $g$.}
	\label{fig:diag_neg}
\end{figure}

A priori, the diagonal decoupling can have all four terms in Eq.~\eqref{eq:O}, while satisfying the relationship~\eqref{eq:diag-}. We have computed the general dispersion relation for the diagonal decoupling, and minimized the corresponding ground-state energy over these four parameters. We found that the minimum occurs at $A=C=D=0$ for all value of $g$, while there is a phase transition at strong enough interaction, at which only $B$ becomes nonzero. The results of this numerical minimization are shown in Fig.~(\ref{fig:diag_neg}).

\begin{figure}[]
	\includegraphics[width=0.5\columnwidth]{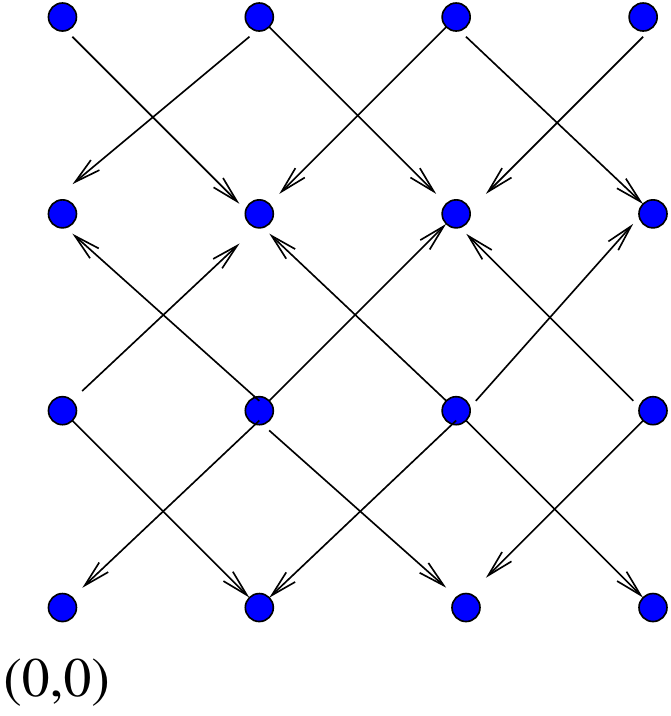}
	\caption{Diagonal ground state for $g<0$. An 
arrow pointing along a diagonal from site $\vec r_1$ to site $\vec r_2$ indicates that 
$i\langle \gamma_{\vec r_1}\gamma_{\vec r_2}\rangle >0$.}
	\label{DgN}
\end{figure}

As discussed in the next section, this is indeed expected from the low energy field theory. To simplify the analytical discussion, we thus study the lattice model for the following diagonal candidate mean-field ground states:
\be \bra i\gamma_{m,n}\gamma_{m+1,n+1}\ket=\bra i\gamma_{m+1,n}\gamma_{m,n+1}\ket=B(-1)^{n},\label{DGON}\ee
which is illustrated in Fig. (\ref{DgN}). As seen in Fig. (\ref{DgN}), the diagonal order of Eq. (\ref{DGON}) simply corresponds to  a nonzero diagonal hopping $t_2$ [see Eq.~\eqref{eq:t2}]. This order parameter breaks time reversal and parity symmetry while preserving the other symmetries.

The mean-field interaction Hamiltonian can then be written as 
%
\be H^d_{int}\approx (igB/8)\sum_{m,ns,s'}\gamma_{m,2n}\gamma_{m+s,2n+s'}-2gWLB^2/64,\ee
where $s$ and $s'$ are summed over $\pm 1$. 
The dispersion relationship then becomes
\be E_\pm =\pm 4\sqrt{(t\sin k_x)^2+(t\sin k_y)^2+(2gB\cos k_x\cos k_y)^2}.\label{EODN}\ee
In terms of the above relationship [Eq.~\eqref{EODN}], the ground-state energy per unit area is given by
\be {\cal \epsilon}\approx -{gB^2}-{1\over (2\pi )^2}\int_0^\pi dk_x\int_{-\pi /2}^{\pi /2}dk_yE_+(\vec{k}).
\ee
Treating $gB$ as a small parameter and Taylor expanding, we see that the transition to $B\neq 0$ occurs at
\be {1\over g_c}\approx-{2\over \pi^2t}\int d^2k{(\cos k_x\cos k_y)^2\over \sqrt{(\sin k_x)^2+(\sin k_y)^2}}.\ee

%

Comparing Figs.~(\ref{ss}) and (\ref{fig:diag_neg}), we observe that the phase transition in the diagonal case Fig.~(\ref{fig:diag_neg}) occurs at a smaller value of $|g|\approx 0.9$ than the transition in the vertical case   Fig.~(\ref{ss}). However, the energies of the vertical mean-field states are lower than the energy of the diagonal state. We thus find 3 phase transitions at $g<0$. For $-0.9<g<0$, all order parameters vanish 
and the system is in the gapless phase. For $-1.4<g<-0.9$, diagonal order occurs. For $g<-1.4$ vertical order occurs, with only $A$ nonzero for $-2.4<g<-1.4$ and both 
$A$ and $B$ nonzero for $g<-2.4$.

Topological analysis of the phases for $g>-0.9$ is analogous to the $g=0$ case. (The mean-field Hamiltonian is not modified). In particular, adding small $t_2$ to the gapless phase $-0.9<g<0$ makes it gapped and having Chern number $\mathcal{C} = - \mathrm{sgn}\;(t_2/t)$.

\section{Field theory approach to interacting model: nature of the phase transitions}
\label{sec:ft}
The gapless phases for $g_{c3}<g<g_{c4}$ have a low energy field theory description with emergent Lorentz and U(1) symmetries. 
The transitions out of these phases are both expected to be second order and therefore in the same universality 
classes as critical points in corresponding Lorentz invariant field theories. By contrast, the transitions at 
$g_{c2}$ and $g_{c1}$ separate gapped phases and appear to be first order. Therefore, the emergent symmetries and field 
theory description do not apply. In this section, we analyze the transitions at $g_{c4}$ and $g_{c3}$ using known 
results on Lorentz invariant field theories. The broken symmetry phases for $g<g_{c2}$ can also be understood from a 
field theory perspective, which we provide here. However, this field theory perspective is only expected to be of relevance 
if a continuous transition occurs, which does not seem to be the case. We also analyze, from a field theory perspective, 
the candidate diagonal broken symmetry state for $g>0$, although it did not occur in the mean-field theory analysis of 
the previous section. 

\subsection{Second-order transitions}
For  $0<g<g_{c4}$, the mean-field  approach of the previous section predicted broken rotational symmetry, with the hopping parameter effectively stronger in the $y$ (or $x$) direction. In the 
field theory we may simply rescale the $y$-coordinate by a factor of 
$(1-gA/t)$
 to recover full Lorentz invariance. We note that such rescaled rotational invariance is quite common 
in critical phenomena; it occurs, for example in the two-dimensional classical Ising model at the critical temperature with different couplings in $x$ and $y$ directions. 

The prediction of a  transition at $g=0$ from a phase with unbroken symmetry at $g<0$ into a phase with effectively stronger hopping in the vertical (or horizontal) direction 
at $g>0$ is somewhat surprising, given that the interactions are irrelevant.  However, since this transition is predicted to be first order, the scaling dimension of 
the interaction term does not play a role, so this appears possible. 

The interaction term in the field theory approximation can be written in two ways:
\be {\cal H}_{int}=-64g\psi^\dagger_1\psi^\dagger_2\psi_2\psi_1=32g(\bar \psi \psi )^2.\ee
For $g>0$ this can be exactly rewritten in terms of a complex scalar field $\phi$ and for 
$g<0$ in terms of a real scalar field $\sigma$:
\bea {\cal H}_{int}&\to & m^2|\phi |^2+8m\sqrt{g}(\psi^\dagger_1\psi^\dagger_2\phi +{\rm H.c.})\ \  (g>0)\nonumber \\
&\to& {m^2\over 2}\sigma^2+8m\sqrt{-g}\bar \psi \psi \sigma \ \  (g<0),\eea
through a Hubbard-Stratonovich transformation. This is related to the mean-field factorization used in Sec. (\ref{MFT_PD}).        
We can promote the fields $\phi$ and $\sigma$ to relativistic massive fields, with real-time Lagrangians:
\bea {\cal L}&=&\bar \psi i\gamma^\mu\partial_\mu \psi +|\partial_\mu \phi |^2-m^2|\phi |^2
+g_1(\psi^\dagger_1\psi^\dagger_2\phi +h.c.)\nonumber \\
&-&g_2|\phi |^4,\ \  (g>0)\nonumber \\
&=&\bar \psi i\gamma^\mu\partial_\mu \psi+{1\over 2}(\partial_\mu \sigma )^2-{m^2\over 2}\sigma^2+
g_1\bar \psi \psi \sigma -2g_2\sigma^4.\nonumber \\
&&\ \  (g<0).\label{LFB}\eea
Here
\bea 64g&=&{g_1^2\over m^2}, (g>0)\nonumber \\
  &=&-{g_1^2\over m^2}, (g<0).\eea
  We have set the velocities to 1 for both fermion and boson fields. 
As long as   $m^2$ is positive and  large enough we get the correct low energy theory. 
Note that 
reducting $m^2$ for fixed $g_1$ corresponds to increasing $|g|$.
We can think of this transition as being driven by reducing $m^2$ and letting it change sign.  We expect the transitions in this fermion-boson models 
to be in the same universality classes as the transitions in our pure fermion model. The first 
Lagrangian in Eq. (\ref{LFB}) is the one studied in Refs. [\onlinecite{Thomas2005,Lee2007,Zerf2016,Klebanov2016}]. It is expected to have a transition into a superfluid 
phase.   This transition is believed to be SUSY. Both 
$\phi$ and $\psi$ are massless at the critical point, forming a supermultiplet. 
The second Lagrangian in Eq. (\ref{LFB}) is discussed in Ref. [\onlinecite{Klebanov2016}]. These authors refer to it as the Gross-Neveu-Yukawa model. It also has a transition to a broken symmetry phase 
as we reduce $m^2$ and take it negative. In this case the broken symmetry is just $Z_2$, $\sigma \to -\sigma$, 
$\bar \psi \psi \to -\bar \psi \psi$. 
They study the critical exponents at this transition using $\epsilon$-expansion and other techniques. 
It is not the same as the ordinary Ising transition, which occurs when $g_1=0$, due to the presence of 
the massless fermion field. On the other hand, the transition is not SUSY.

Now we demonstrate the equivalence of these transitions with the ones in the lattice model at $g_{c4}$ and $g_{c3}$.
The order parameter for the transition at $g_{c4}$ is
\be \bra i\gamma_{m,n}\gamma_{m,n+1}\ket=B(-1)^n\ee
or, equivalently,
\be \bra i\gamma^e_{2n,m}\gamma^o_{2n,m}\ket= \bra i\gamma^e_{2n,m}\gamma^o_{2n-2,m}\ket=B\ee
for real $B$ with two different states depending on the sign of $B$.
This implies, in the low energy limit,
\be 8i\bra\chi^{e+}\chi^{o+}+\chi^{e-}\chi^{o-}\ket=B\ee
The other two ground states, related by a $\pi /2$ rotation, have 
\be i\bra\gamma_{m,n}\gamma_{m+1,n}\ket=B(-1)^{m+n}\ee
or, equivalently, 
\be i\bra\gamma^e_{m,2n}\gamma^e_{m+1,2n}\ket=-i\bra\gamma^{o}_{m,2n}\gamma^o_{m+1,2n}\ket=(-1)^mB,\ee
which, using Eq. (\ref{chid}), implies
\be 16i\bra\chi^{e+}\chi^{e-}\ket=-16i\bra\chi^{o+}\chi^{o-}\ket=B.\ee
Note that the Dirac ferimions, defined in Eq. (\ref{Dirac}), obey
\be \psi_1\psi_2 = (\chi^{e+}\chi^{o+}+\chi^{e-}\chi^{o-})+i(\chi^{e+}\chi^{e-}-\chi^{o+}\chi^{o-}).
\ee
Thus, vertical order and horizontal order imply respectively
\bea \bra\psi_1\psi_2\ket&=&-iB/8\ \  (\hbox{vertical})\nonumber \\
&=&B/8\ \  (\hbox{horizontal}).\eea
Recall that the low energy field theory has an emergent $U(1)$ symmetry, corresponding to conservation of Dirac fermion number. The broken symmetry states correspond to 
spontaneous breaking of this emergent $U(1)$ symmetry. More general ground states would have an arbitrary phase for $\bra \psi_1\psi_2\ket $, corresponding to a linear combination 
of the vertical and horizontal order. 
\be E\approx \sqrt{(v\vec k)^2+(8gB)^2}\ee
after rescaling the $y$-coordinate.  The mean-field transition can also be obtained in the field theory and is again predicted to be second order. The second order nature of the 
transition, in mean-field theory, is a result of a cubic term in $|B|$ in the Taylor expansion of the ground-state energy, of Eq. (\ref{EDel}), which results from the small $k$ region. 
This cubic term can be obtained in the relativistic approximation using
\be {E_0/2WL}\approx gB^2-{1\over 8\pi^2}\int_{k<\Lambda}d^2k\sqrt{(vk)^2+(8gB)^2}.\ee
At small $B$ this gives a universal cubic term $(1/6\pi )|8gB|^3$, independent of the cut-off $\Lambda$.  The terms in the effective Hamiltonian which break the $U(1)$ 
(and Lorentz) symmetry are of dimension 6 or higher, more irrelevant than the relativistic interaction term. It is therefore plausible that they remain irrelevant at the 
critical point. 
A novel feature of our model is that the  boson, and hence the supersymmetry, 
results from a bosonic bound state of the fermions. This is very reminiscent of the supersymmetric phase transition discovered in the one-dimensional version of this model~[\onlinecite{Rahmani2015, Rahmani2015a}]. 
(A field theory discussion of purely fermionic models 
appears in Ref. [\onlinecite{Klebanov2016}].)
Of course, the other novel feature, not occurring in 1D and rather unique to this model, is that the $U(1)$ symmetry, which is spontaneously generated is an emergent symmetry.

Now we consider the transition at $g_{c3}$, 
the first transition, as $g$ is decreased from $0$, which is into the diagonal phase. Equation (\ref{DGON}) implies:
 \be 8i\bra \chi^{e+}\chi^{o+}-\chi^{e-}\chi^{o+}\ket =B.\ee
 In terms of the Dirac fermions defined in Eq. (\ref{Dirac}) this becomes:
 \be 4\bar \psi \psi =-B.\ee
 This corresponds to spontaneous generation of a mass term for the relativistic fermions, spontaneously breaking time reversal and parity.  The dispersion relation of Eq. (\ref{EODN}) in the 
 low energy approximation is
 \be E_\pm =\pm \sqrt{(vk)^2+(2gB)^2},\ee
 corresponding to a mass term.  Again the transition is second order due to a term cubic in $|B|$ in the mean-field ground-state energy.

\subsection{Other order parameters}

The diagonal order parameter for $g>0$, which apparently does not become nonzero, corresponds, from Eq. (\ref{DG0}), to
\be i\bra \gamma^e_{m,2n}\gamma^o_{m+1,2n}\ket =A(-1)^m.\ee
In the field theory, this corresponds to
\be 8i\bra \chi^{e-}\chi^{o+}-\chi^{e+}\chi^{o-}\ket =A.\ee
Noting that
\be  \psi^\dagger \psi =2i(\chi^{e-}\chi^{o+}-\chi^{e+}\chi^{o-}),\ee
we see this corresponds to 
\be \bra \psi^\dagger \psi \ket =A/4.\ee
This order parameter breaks charge conjugation symmetry and corresponds to adding a chemical potential coupled to the emergent conserved charge.  
The energy bands of Eq. (\ref{Eposdiag}) correspond, at low energies, to energies $vk-8A$ for particles and $vk+8A$ for holes. 

 Next, we
 consider the staggered vertical order of Eq. (\ref{VAF}), which we found to occur for $g<-1.4$. Setting $D=0$, this corresponds to
 \be i\bra \gamma^e_{m,2n}\gamma^o_{m,2n}\ket =i\bra \gamma^e_{m,2n}\gamma^o_{m,2n-2}\rangle=C(-1)^m.\ee
 In the continuum limit this implies
 \be 8i\bra \chi^{e+}\chi^{o-}+\chi^{e-}\chi^{o-}\ket =C.\ee
In terms of the Dirac fermions of Eq. (\ref{Dirac}), this becomes:
\be \bra \bar \psi \gamma^1\psi \ket =-4C.\label{current}\ee

There is an equivalent mean-field ground state, rotated by $\pi /2$ (uniform horizontal order), with order parameter
\be i\bra \gamma_{m,n}\gamma_{m+1,n}\ket =C(-1)^m.\ee
This implies
\be 16i\bra \chi^{e+}\chi^{e-}\ket = 16i\bra \chi^{o+}\chi^{o-}\ket =C.\ee
Noting that
\be \bar \psi \gamma^2\psi =-2i(\chi^{e+}\chi^{e-}+\chi^{o+}\chi^{o-}),\ee
we see that this equivalent state has 
\be\bra  \bar \psi \gamma^2\psi \ket =-4C.\ee
As expected, this is obtained by a $\pi /2$ rotation of the state of Eq. (\ref{current}). Noting that $\bar \psi \vec \gamma \psi =\vec J$, the current operator, we see that 
these states have a spontaneously generated current, flowing in the $x$ or $y$ direction. More generally, in the field theory, a linear combination of these states would 
have the same energy, corresponding to the current flowing in an arbitrary direction in the $x$-$y$ plane. 
The dispersion relation, for order in the $y$ direction, at small $k$ becomes
\be E=(vk_x+ 8Cg)^2+(vk_y)^2.\ee
(Note that in the previous section, $k_x$ was restricted to positive values.  Interpreting the second solution as corresponding to $k_x<0$ we have a fixed sign for $A$.) 
The corresponding mean-field Hamiltonian density is
\be {\cal H}=-gC^2+\bar \psi [-\gamma^1(iv\partial_1-8gC)-\gamma^2iv\partial_2]\psi .\ee
We see that $C$ corresponds to a vector potential in the $x$ direction, which leads to a nonzero current. Based on the mean-field calculations in the previous section,  we 
expect the  transition into this phase with a spontaneously generated current to be first order.

\subsection{Including second-neighbor hopping term}
As discussed in Sec. (\ref{symm}), the second-neighbor hopping term, $\propto t_2$, breaks time reversal symmetry and parity (spatial reflection) symmetry.  As shown in Sec. (\ref{NIS}), it produces an excitation gap which, 
as shown in Sec. (\ref{FTS}), corresponds to a mass term in the low energy effective field theory.  While  time reversal and parity are broken by this mass term, Lorentz invariant, 
the emergent $U(1)$ symmetry and charge conjugation (produce of parity and time reversal) remain good emergent symmetries. 

We still expect the emergent superfluid phase to occur for sufficiently large $g>0$, as shown in Sec. (\ref{MFT_PD});
see Fig. (\ref{f2}).   However, since the fermion now has a finite mass at the critical point, whereas the Goldstone boson is 
massless, we expect that this transition will no longer  be supersymmetric, but instead fall into the usual (2+1) dimensional $U(1)$ breaking universal class.  The first transition transition
at negative $g$, at $g_{c3}$, no longer occurs since it corresponds to spontaneous breaking of time reversal and parity which are already explicitly broken by the mass term. 
On the other hand, the  first-order transitions at $g_{c1}, g_{c2}<0$ into  phases with broken spatial rotation symmetry, corresponding to a current flowing in an arbitrary direction, may still occur. 

\section{Conclusions}
\label{sec:sum}
We have studied one of the simplest possible 2D models of interacting Majorana modes, using a combination of mean-field theory and renormalization group methods.  We 
find 6 different phases as the coupling strength is varied over both positive and negative values. In particular there are gapless phases at sufficiently weak coupling with emergent
Lorentz invariance and U(1) particle number conservation.  These are separated by continuous transitions from an emergent superfluid phase for attractive interactions and from a phase with 
broken time reversal for repulsive interactions.  The superfluid transition is predicted to exhibit supersymmetry.  The model does {\it not} appear to be amenable to 
Monte Carlo methods since the Majorana modes are not doubled; we intend to present density-matrix renormalization group results on ladders in a later paper.

\acknowledgements
The authors would like to thank Marcel Franz, Tim Hsieh, Igor Klebanov, Joseph Maciejko and Kyle Wamer  for very helpful discussions and emails. This research 
was supported in part by NSERC Discovery Grant 04033-2016 (IA) and the Canadian Institute for Advanced Research (IA). DIP is grateful to KITP, where part of the research was discussed with the support of NSF PHY11-25915.

\appendix
\section{$\pi /2$ rotation symmetry of lattice model}
Under the transformation of Eqs. (\ref{po2}) and (\ref{s}), the nearest-neighbor hopping term transforms as:
\bea &&\sum_{m,n}\gamma_{m,n}[(-1)^n\gamma_{m+1,n}+\gamma_{m,n+1}]\nonumber \\
&&\to \sum_{m,n}\gamma_{-n,m}[\gamma_{-n,m+1}-(-1)^m\gamma_{-n-1,m}]
.\eea
Now redefining the summation variables by $n\to -n$ then $m\leftrightarrow n$, and switching the order of 
the $\gamma$'s in the second term, we recover the original expression. 
The interaction term transforms as:
\bea &&\sum_{m,n}\gamma_{m,n+1}\gamma_{m+1,n+1}\gamma_{m+1,n}\gamma_{m,n}\nonumber \\
&&\to 
-\sum_{-n-1,m}\gamma_{-n-1,m}\gamma_{-n-1,m+1}\gamma_{-n,m+1}\gamma_{-n,m}.\eea
Redefining the summation variables by $n\to -n$ then $m\leftrightarrow n$, this becomes:
\bea && -\sum_{m,n}\gamma_{m-1,n}\gamma_{m-1,n+1}\gamma_{m,n+1}\gamma_{m,n}\nonumber \\
&&=\sum_{m,n}\gamma_{m-1,n+1}\gamma_{m,n+1}\gamma_{m,n}\gamma_{m-1,n}
.\eea
Finally, shifting the summation variable, $m\to m+1$ we recover the original term. 
The second-neighbor hopping term transforms as
\bea && \sum_{m,n,s,s'}\gamma_{m,2n}\gamma_{m+s,2n+s'}\nonumber \\
&& \to \sum_{m,n,s,s'}(-1)^m\gamma_{-2n,m}\gamma_{-2n-s',m+s},\eea
where $s$ and $s'$ are summed over $\pm 1$. Redefining summation variables by $n\to -n$ then $m\leftrightarrow n$ 
together with $s'\to -s$, $s\to s'$, this becomes
\be \sum_{m,n,s,s'}(-1)^n\gamma_{2m,n}\gamma_{2m+s,n+s'}.\ee
Now treating the even and odd $n$ terms separately, this can be written:
\be \sum_{m,n,s,s'}[\gamma_{2m,2n}\gamma_{2m+s,2n+s'}-\gamma_{2m,2n+1}\gamma_{2m+s,2n+1+s'}].\ee
In the second term we then  switch order of the  $\gamma$ factors, redefine $ 2n+1+s'\to 2n$ (for each value of $s'$)
and redefine $2m+s\to 2m+1$ (for each value of $s$). The term then becomes
\be \sum_{m,n,s,s'}[\gamma_{2m,2n}\gamma_{2m+s,2n+s'}+\gamma_{2m+1,2n}\gamma_{2m+1+s,2n+s'}].\ee
Combining the two terms gives the original expression.

\section{Normalization in Eq. (\ref{chid})}
Equation (\ref{chid}) implies, for example,  that $\chi^{e+}$ contains the Fourier modes of $\chi^e$ near $\vec k=0$, divided by $2\sqrt{2}$. Thus, also using Eq. (\ref{FT}), 
\be \chi^{e+}={1\over 2\sqrt{2}}\sqrt{2\over WL}\sum_{|\vec k|<\Lambda}e^{i\vec k\cdot \vec r}\gamma^e_{\vec k}.\ee
From Eq. (\ref{k}), we see that $\Delta k_x=2\pi /L$ and $\Delta k_y=\pi /W$. Thus
\bea \{ \chi^{e+}(\vec r),\chi^{e+}(\vec r')\}&=&{1\over 8}{2\over WL}\sum_{k<\Lambda }e^{i\vec k\cdot (\vec r-\vec r')}
\nonumber \\
&=&{1\over 2}\int_{k<\Lambda }{d^2k\over (2\pi )^2}e^{i\vec k\cdot (\vec r-\vec r')}\nonumber \\
&\approx& {1\over 2}\delta^2(\vec r-\vec r'),
\eea
showing that $\chi^{e+}$ has the correct normalization. A similar result holds for all 4 fields, $\chi^{e/o\pm}$.

\bibliography{M2D.bib}

\end{document}